\documentclass[aps,prl,twocolumn,groupedaddress,nofootinbib]{revtex4-1}
\usepackage{amsmath}
\usepackage{graphicx}
\usepackage{sistyle}
\usepackage{bm}
\usepackage{natbib}
\usepackage{dsfont}
\usepackage{epsfig}
\usepackage{feynmf}
\usepackage{blindtext, rotating}
\usepackage{mathtools}
\usepackage{dsfont}
\usepackage{subcaption}
\usepackage{physics}
\usepackage{amsfonts}
\usepackage{ragged2e}
\usepackage{tcolorbox}
\usepackage{color,soul}

\DeclareCaptionJustification{justified}{\justifying}

\begin{document}
\title{Molecular machines for quantum error correction}
% Quantum error correcting machines
% Quantum Error Correcting Molecular Machines
% Quantum error correcting robots
% Active quantum error correcting machines
\author{Thiago Guerreiro}
\email{barbosa@puc-rio.br}

\affiliation{Department of Physics, Pontif\'icia Universidade Cat\'olica, Rio de Janeiro, Brazil}

\begin{abstract}
%Quantum entanglement finally comes to life!
Inspired by biological molecular machines we explore the idea of an active quantum robot whose purpose is delaying decoherence. 
A conceptual model capable of partially protecting arbitrary logical qubit states against single physical qubit errors is presented. Implementation of an instance of that model - the entanglement qubot - is proposed using laser-dressed Rydberg atoms. Dynamics of the system is studied using stochastic wavefunction methods.
\end{abstract}

%\keywords{Entanglement, dispersive optomechanics}

\maketitle

\section{Introduction}

The living cell can be seen as a Brownian computer \cite{Bennett1982}. At its core, machines of molecular dimensions store, correct and process information in the presence of noise, with the goal of keeping the state of the living creature away from thermodynamical equilibrium. 
The machinery of life \cite{Goodsell1993} is responsible for gene expression, matter transport across the cell and energy harvesting, among a vast number of other tasks \cite{Alberts}. 
%In parallel to developments in nanotechnology, we are now entering the age of quantum information processing, which holds great expectations for quantum chemistry simulations and biology. In this work, we invite the exploration of the 
An example of such molecular devices is \textit{RNA polymerase} (RNAP): an enzyme with $\sim 40.000$ atoms, roughly $ \SI{10}{nm} $ of linear size, capable of synthesising a strand of RNA from a DNA template in the presence of Brownian noise, at error rates as low as $ 10^{-7} $ \cite{Milo}. 
Molecular devices such as RNAP have inspired nanotechnology \cite{Feynman, Zhang2018} and various artificial molecular machines were built, such as molecular ratchets \cite{Serreli2007}, pumps \cite{Stoddart2015}, motors \cite{Kassem2017}, and gene editing tools \cite{CRISPR}. 

Detailed unified understanding of biological molecular machines according to the tradition of theoretical physics is yet to be achieved \cite{Bialek}, but there is little doubt that experimental \cite{Block} and computational methods \cite{Bressloff} in physics play a key role in that endeavour.
It is also expected that the coming age of quantum information processing will illuminate biological systems through simulation of quantum chemistry \cite{Google_ai} and quantum enhanced learning \cite{Outeiral2020, Emani2021}. 
Conversely \cite{Frauenfelder2014}, one could ask whether biological molecular machines will inspire new ideas for engineering autonomous molecular-sized quantum information processing devices with the goal of keeping quantum states away from thermodynamical equilibrium. It is the purpose of this work to explore this idea. 

\begin{figure}[ht!] % [width=8.6 cm]
%    \centering
    \includegraphics[width =0.5\textwidth]{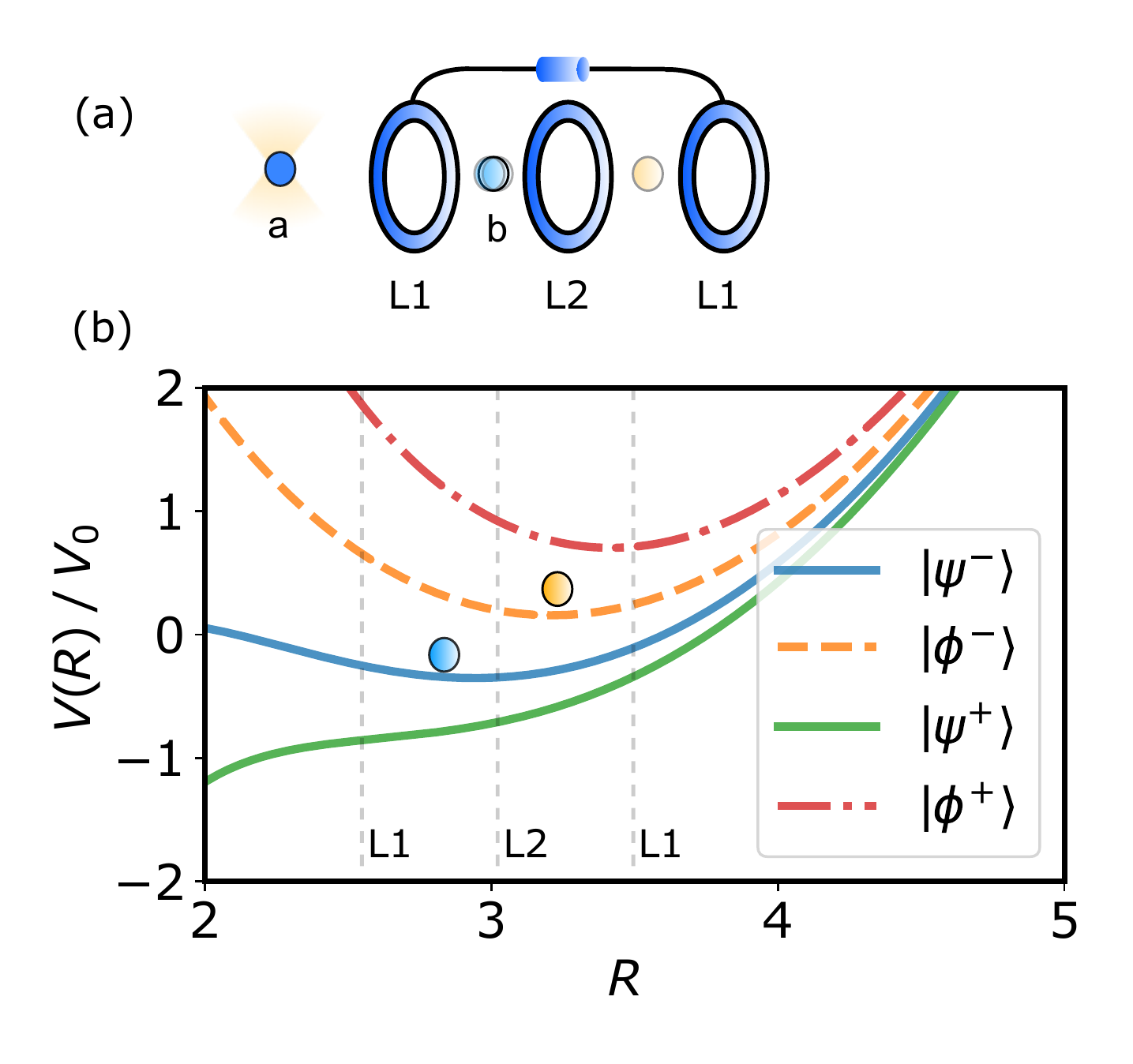}
    \caption[]{(a) Schematics for a conceptual qubot model capable of partially protecting an arbitrary logical qubit state against decoherence. (b) Example of a possible potential landscape describing the interaction between the nucleus atoms; for this plot the radial dependence of \eqref{interaction} is considered $ J_{\alpha}(R) = ( d^{2} / R^{3} ) j_{\alpha} $, with $j_{y} = -3 j_{x} , j_{z} = 6 j_{x} $. }\label{landscape}    \label{color}
\end{figure}
%$V_{0} = 0.4 k^{3}d^{2} $

A quantum molecular machine would be a device composed of at most a few thousand atoms capable of autonomously storing, protecting and/or processing quantum states in the presence of external decoherence and thermalization. We refer to these bio-inspired devices as quantum robots, or qubots \cite{Guerreiro2020}. 
Devising qubots is a problem in coherent quantum chemistry \cite{Carr2009, Krems} much like engineering artificial molecular machines is a problem in synthetic chemistry \cite{Lau2017}. Hence, the ultracold atom \cite{Budker} and molecular toolbox \cite{Ospelkaus2010, Liu2017} is expected to play a key role in the conception of these active quantum devices. 
As we will see, qubots exploit open system dynamics to achieve their purpose and thus have a close connection to the idea of engineered environments constructed to produce desired quantum states \cite{Plenio1999, Plenio2001, Diehl2008, Verstraete2009, Vacanti2009, Reiter2012, Brask2015, Reiter2017}. Their nature, however, is much closer to that of artificial molecular ratchets and pumps that respond to the environment and consume resources to maintain nonequilibrium states \cite{Cheng2015}. 

In what follows, we explore various aspects around the idea of qubots.  
%Our exploration of qubots proceeds as following. 
We begin by introducing a conceptual model for a quantum robot capable of partially protecting a logical qubit state against single physical qubit errors. It is interesting that the model can handle almost all combinations of phase and bit-flip errors since, as pointed out by Kitaev, \textit{it is generally easy to get rid of one kind of errors, but not both} \cite{Kitaev2001}. The construction is somewhat inspired by the surface code \cite{Fowler2012}, only here syndrome detection and correction are part of the system's dynamics rather than a consequence of measurement followed by external conditional action.
Next, a specific physical implementation of instances of the model based on laser-dressed Rydberg atoms is discussed. More specifically, we exhibit potential landscapes implementing an \textit{entanglement qubot}, a device that stabilizes a Bell state against single qubit errors. The stabilized Bell state is only one possible state of the logical qubit, but in this case we can view the qubot as preserving a maximally entangled state. An ensemble of entanglement qubots could therefore preserve vast amounts of entanglement, a useful resource.
Simulation of the entanglement qubot dynamics is performed with the help of stochastic wavefunction methods, and we evaluate the effects of coupling the motional degrees of freedom of the robot to an external heat bath. 
We conclude with a discussion on potential future developments regarding active quantum matter.

\section{Conceptual model}

We would like to introduce the conceptual model of a quantum robot capable of protecting an arbitrary logical qubit state against errors.
Our quantum robot consists of two parts, called the \textit{nucleus} and the \textit{correctors}. See Figure \ref{landscape}(a) for a schematic representation. 
A pair of particles denoted $ a $ and $ b $ constitute the nucleus.
%The particles have mass $ m $ and are trapped in a one-dimensional potential. 
Quantum information is stored in the particles' internal spin degrees of freedom taken to be two spin 1/2 systems with Hilbert space $ \mathbb{C}^{2} \otimes \mathbb{C}^{2}   $ and basis states denoted $ \lbrace  \vert 0 \rangle  \vert 0 \rangle ,  \vert 0 \rangle  \vert 1 \rangle,  \vert 1 \rangle  \vert 0 \rangle ,  \vert 1 \rangle  \vert 1 \rangle \rbrace $.

Particle $ a $ is held fixed at the origin by an optical tweezer while $ b $ is subject to the potential 
\begin{eqnarray}
V(R) = V_{t}(R) + V_{I}(R) \ ,
\end{eqnarray}
where $ R $ is the relative distance between $ a $ and $ b $, $ V_{t}(R) $ is a trap potential for particle $ b $ and
\begin{align}
V_{I}(R) =  J_{z} Z_{a} Z_{b} + J_{x} X_{a} X_{b} + J_{y} Y_{a} Y_{b}  \ ,
\label{interaction}
\end{align}
is the interaction energy between the particles, where $ X_{\lambda}, Y_{\lambda}, Z_{\lambda}  $ are the Pauli operators for particle $ \lambda $ ($ = a, b $) and the coefficients $ J_{\alpha} = J_{\alpha}(R)  $ form a spatial-dependent spin-spin interaction pattern.  We assume for simplicity that particle $ b $ can only move along the direction $ \hat{R} $.

As an example of trap potential one may consider an optical tweezer,
\begin{eqnarray}
V_{t}(R) &=& V_{0}  \left( R - \delta \right)^{2} \ .
\label{lattice}
\end{eqnarray}
where $ V_{0} $ and $ \delta $ are constants. 
%Optical tweezers also provide an interesting option for $ V_{t}(R) $ which shall be explored later.
Tunneling outside the confining potential is considered negligible.
Note also that dipole-dipole interactions among atoms and polar molecules is of the form \eqref{interaction}, and typically for molecules \cite{Wei2011, Pietraszewicz2013} and spin impurities in diamond \cite{Choi2017},
\begin{eqnarray}
 J_{\alpha} = (d^{2} / R^{3} ) j_{\alpha} \ , 
 \label{pattern}
\end{eqnarray} 
where $ d $ is the dipole moment \cite{Weinberg2012} and $ j_{\alpha} $ a proportionality constant with $ \alpha = x, y, z $. Through the remaining of this section we will consider this radial dependence as an illustration of the qubot functioning. Note however that effective spin interactions of the so-called $XYZ$ form with more general radial dependencies can be engineered within a number of different systems, including trapped ions \cite{Porras2004, Kim2010}, atoms in dressed Rydberg states \cite{Glaetzle2015, Bijnen2015} and microwave-excited polar molecules in optical lattices \cite{Micheli2006, Brennen2007}. 
In the next section an implementation using laser dressed Rydberg atoms will be discussed.

Bell states of the particles' spins are eigenstates of $ V_{I}  $ with eigenvalues given by
\begin{eqnarray}
V_{I} \vert \psi^{-} \rangle &=& \left(  -J_{x} - J_{y} - J_{z}   \right) \vert \psi^{-} \rangle \label{V_1} \ , \\
V_{I} \vert \phi^{-} \rangle &=& \left(  -J_{x} + J_{y} + J_{z}   \right) \vert \phi^{-} \rangle \label{V_2} \ , \\
V_{I} \vert \psi^{+} \rangle &=& \left(  J_{x} + J_{y} - J_{z}   \right) \vert \psi^{+} \rangle \label{V_3}  \ , \\
V_{I} \vert \phi^{+} \rangle &=& \left(  J_{x} - J_{y} + J_{z}   \right) \vert \phi^{+} \rangle \label{V_4}  \ .
\end{eqnarray}
This implies that the total potential $ V(R) $ exhibits collective spin-dependent landscapes. 

As an example consider the trap potential \eqref{lattice} and the spin pattern \eqref{pattern}. If local equilibrium positions $ R_{0}(\vert \psi \rangle) $ exist, they satisfy the condition
\begin{eqnarray}
R_{0}^{4} (R_{0} - \delta) =  \dfrac{ 3d^{2} \langle \psi \vert   W \vert \psi \rangle}{2 V_{0}}  \ ,
\label{equilibria}
\end{eqnarray}
where $ \langle \psi \vert   W \vert \psi \rangle = \langle \psi \vert   \left(  j_{z} Z_{a} Z_{b} + j_{x} X_{a} X_{b} + j_{y} Y_{a} Y_{b} \right) \vert \psi \rangle $ are possible expectation values with respect to each of the four Bell states. 
Figure \ref{landscape}(b) shows the total potential landscape seen by particle $ b $ for each of the spin Bell states, displaying the spin-dependent potentials. Note that the state $ \vert \psi^{+} \rangle $ does not exhibit a minimum; this is not a problem provided the protected logical qubit states do not involve $ \vert \psi^{+} \rangle $.

In between equilibrium points of the potential landscapes in Figure \ref{landscape}(b) there are \textit{corrective} sites, where devices we call correctors are present. Correctors are represented in Figure \ref{landscape}(a) as \textit{loops}. The function of the corrective devices is executing a unitary operation on the spin subspace once the particle approaches their site. 
There are two correctors, denoted $ L1$ and $ L2 $.
For illustration of the device functioning, in the remaining of this section we treat the correctors $ L1 $ and $ L2 $ as qubits.
Note however that there are a number of ways of implementing such devices and alternatives to the qubit model will be discussed in the following implementation section.
%\footnote{It is important to note that such devices need not be qubits, and further details on their physical implementation is discussed in the following sections.}.

Consider the $ L1 $ device has basis states $ \lbrace \vert \mu_{0}^{1} \rangle, \vert \mu_{1}^{1} \rangle \rbrace $. Whenever the particle enters one of the $ L1 $ loops, the unitary operation $ Z_{b}  X_{L1}$ is executed, where  $ X_{L1}  = \vert \mu_{0}^{1} \rangle \langle \mu_{1}^{1} \vert + \vert \mu_{1}^{1} \rangle \langle \mu_{0}^{1} \vert $. It is important that $ L1 $ is insensitive to whether particle $ b $ entered the innermost or outermost loop, since obtaining that information would collapse the spin state of the system as it is correlated to motion.
The $ L2 $ system, or middle corrector, has basis states $ \lbrace \vert \mu_{0}^{2} \rangle, \vert \mu_{1}^{2} \rangle \rbrace $ and whenever particle $ b $ enters $L2 $, the unitary $ X_{b}  X_{L2} $ is executed, where $ X_{L2} $ is once again the bit-flip operator on the corresponding basis states of $ L2 $. 

We have the following operations:
\begin{align}
L1: \  Z_{b}  X_{L1} \ , \  \ L2: \ X_{b}  X_{L2} \ .
\label{loop_eqs}
\end{align}
Note these unitaries act on the spins \textit{conditional} on the particle's position. Hence, when tracing out the position degree of freedom, action of the corrective sites manifests as dissipative maps on the spin subspace.

Logical basis states of the nucleus are defined as 
\begin{eqnarray}
\vert \bar{0} \rangle &=&  \vert \psi^{-} \rangle 
\\
\vert \bar{1} \rangle &=&  \vert \phi^{-} \rangle 
\end{eqnarray} 
and an arbitrary logical qubit state is
\begin{eqnarray}
\vert \Psi \rangle = \alpha  \vert \bar{0} \rangle  + \beta \vert \bar{1} \rangle 
\end{eqnarray}
Note that a superposition of the $ \vert \bar{0} \rangle, \vert \bar{1} \rangle $ states implies particle $ b $ is in a superposition of singlet and triplet spin states, implying a superposition of different spatial equilibrium points.

To understand how the qubot delays decoherence and partially protects the logical qubit, one must follow carefully what happens to the particles when a physical error occurs in one of the spins. Single physical qubit errors are assumed to be much more likely than multi-qubit errors \cite{Fowler2012} and the depolarizing channel is considered as decoherence model.
A summary of possible errors and how they act on logical basis states is shown in Table \ref{errors}.

\begin{table}[h!]
\centering
\begin{tabular}{cccc}

\hline \hline
                        Error           & $ \ \  \vert \psi^{-} \rangle $   & \multicolumn{1}{c}{$ \ \ \vert \phi^{-} \rangle $} & Corrected state  \\ \hline 
$ X_{a}  $                           & $ - \vert \phi^{-} \rangle $                     & $ - \vert \psi^{-} \rangle $                 & $ - \alpha  \vert \bar{0} \rangle  - \beta \vert \bar{1} \rangle  $ \\
$ X_{b}  $                           & $ + \vert \phi^{-} \rangle $                     & $ + \vert \psi^{-} \rangle $ &      $ \alpha  \vert \bar{0} \rangle  + \beta \vert \bar{1} \rangle  $               \\
$ Z_{a} $                           & $ + \vert \psi^{+} \rangle $                      & $ + \vert \phi^{+} \rangle  $ &       $ \alpha  \vert \bar{0} \rangle  - \beta \vert \bar{1} \rangle  $               \\
$ Z_{b}  $                           & $ - \vert \psi^{+} \rangle $                      & $ + \vert \phi^{+} \rangle $                     & $ -\alpha  \vert \bar{0} \rangle  - \beta \vert \bar{1} \rangle  $  \\
$ Z_{a} X_{a} $                   & $ - \vert \phi^{+} \rangle $                     & $ - \vert \psi^{+} \rangle $             &    $ \alpha  \vert \bar{0} \rangle  - \beta \vert \bar{1} \rangle  $    \\
$ Z_{b} X_{b}  $ & $ + \vert \phi^{+} \rangle $ &  $ - \vert \psi^{+} \rangle $ & $ -\alpha  \vert \bar{0} \rangle  - \beta \vert \bar{1} \rangle  $ \\ \hline \hline
\end{tabular}
\caption{Effect of physical errors on logical basis states and the final corrected state after action of the qubot. \label{errors}}
\end{table}

\begin{figure}[ht!] % [width=8.6 cm]
%    \centering
    \includegraphics[width =0.45\textwidth]{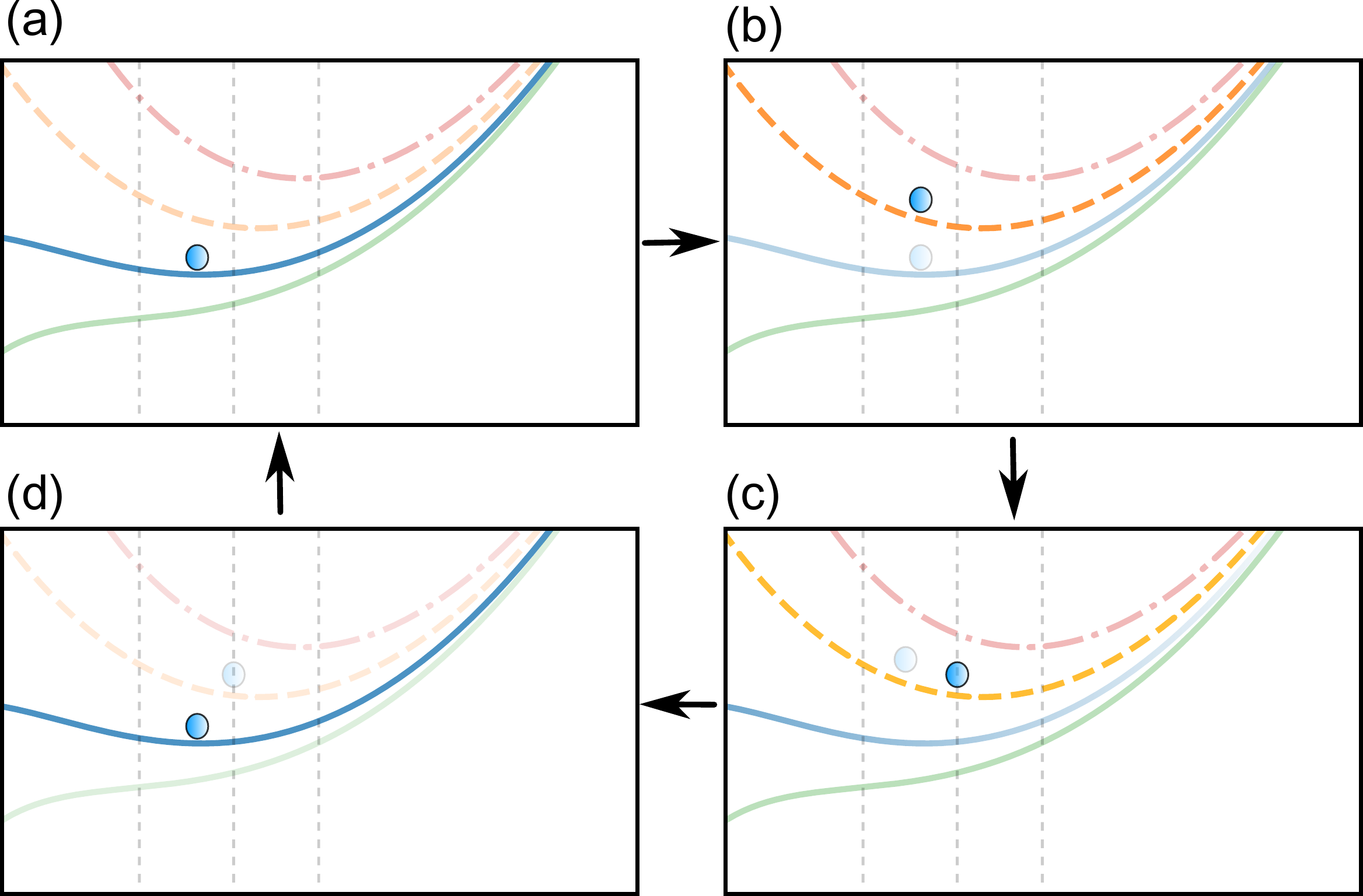}
    \caption[]{Example of a qubot cycle: (a) particle $ b $ rests in its equilibrium position, while the spin states form a singlet $ \vert \bar{0} \rangle = \vert \psi^{-} \rangle $; (b) an error occurs, changing the potential landscape seen by $ b $; (c) the particle is forced into loop $ L2 $, which restores the original spin state and (d) the particle goes back to the original equilibrium position. }\label{cycle}
    \label{color}
\end{figure}

As an illustration, consider the example of a bit-flip in the first spin described by the $ X_{a} $ operator. Initially, an arbitrary logical qubit state $ \vert \Psi \rangle = \alpha  \vert \psi^{-} \rangle  + \beta \vert \phi^{-} \rangle $ is in a superposition of equilibrium positions  $ R_{0}(\vert \psi^{-} \rangle) $ and $ R_{0}(\vert \phi^{-} \rangle) $ given by solutions of \eqref{equilibria}. The $ X_{a} $ error changes the spin state of the particles according to 
\begin{eqnarray}
 \alpha  \vert \psi^{-} \rangle  + \beta \vert \phi^{-} \rangle   \rightarrow - \alpha  \vert \phi^{-} \rangle  - \beta \vert \psi^{-} \rangle
\end{eqnarray}
and hence the particles' interaction potential is changed accordingly. 
After the error, the possible positions of particle $b$ are no longer equilibrium points of the potential landscapes. 
For the case in which $b$ was initially at $ R_{0}( \vert \psi^{-} \rangle) $, the particles repel, forcing $b$ into $L2 $.
Similarly, for $ R_{0}( \vert \phi^{-} \rangle ) $, occurrence of the error causes an attractive force which pulls $ b $ into $ L2 $. 
Once $b$ reaches the loop, the operator $ X_{b} X_{L2} $ is applied, restoring the logical qubit to the original state and driving the system back to the initial superposition of equilibrium points. Naturally this process introduces kinetic energy in the form of phonons, which must be removed if particle $ b $ is to settle back in the original state. This implies the need for a dissipative force acting on $ b $ which could be provided by state-independent cooling of the atom motion. For now, we will assume that such cooling is present, and this \textit{phonon} issue will be discussed further in the implementation section. 

Similar processes occur for $ X_{b} $ and $ Z_{b} $ errors: a combination of spin-motion dynamics and subsequent application of the loop operators corrects errors and restores the system to the initial arbitrary logical state.
The qubot is also able to correct a concatenation of phase and bit-flip errors, given by $ Y_{b} $. Note that this requires a passage through two correctors. 

The present qubot model is not able to correct all errors. As can be seen in Table \ref{errors}, logical basis states transform under $ Z_{a} $ with opposite parity, thus inducing a phase error in the logical qubit. This imparts on the $ Y_{a}  $ error since $ i Y_{a} = Z_{a} X_{a}  $. 
This imperfection can be traced back to the fact that the qubot uses two physical qubits to encode a logical state. 
The quantum Hamming bound \cite{Gottesman1996} implies that for single qubit errors, a minimum of five qubits are required to achieve complete fault tolerance for one logical qubit. 
Despite this partial fault tolerance the qubot can delay decoherence of arbitrary logical qubit states, and for some specific states it is even able to preserve it regardless of the error, as for example the singlet $ \vert \psi \rangle = \vert \psi^{-} \rangle $. 
%Moreover, its simplicity makes it amenable to experimental implementations. 
More general models implementing \textit{perfect} quantum error correcting codes \cite{Laflamme1996} can nevertheless be devised at the expense of more particles or higher spin states.
Note that to protect arbitrary logical qubit states, the qubot potential landscapes must distinguish between all the four elements of the Bell basis, as in Figure \ref{landscape}(b). If the landscape for two or more Bell states is indistinguishable, certain errors will cause no effect upon the atom preventing the action of the correctors. Note also that the order of the potential minima for each Bell state defines the choice of position and action for the corrective sites, as well as the choice of logical basis states. 

It is instructive to consider the qubot operation under a depolarizing channel acting on particle $ b $ alone. Denote environment states as $ \vert e_{j} \rangle $. Decoherence causes the joint particle-environment-corrector state to evolve according to,
\begin{align}
\vert \Psi \rangle \vert e_{0} \rangle \vert \mu_{0}^{1} \mu_{0}^{2} \rangle  \rightarrow \sqrt{1-p} \left( \alpha  \vert \psi^{-} \rangle  + \beta \vert \phi^{-} \rangle  \right) \vert e_{0} \rangle \vert \mu_{0}^{1} \mu_{0}^{2} \rangle \nonumber \\ \nonumber
\\
+ \sqrt{\dfrac{p}{3}} \left( \alpha  \vert \phi^{-} \rangle  + \beta \vert \psi^{-} \rangle  \right) \vert e_{1} \rangle \vert \mu_{0}^{1} \mu_{0}^{2} \rangle \nonumber \\ \nonumber
\\
+  \sqrt{\dfrac{p}{3}} \left( - \alpha  \vert \psi^{+} \rangle  + \beta \vert \phi^{+} \rangle  \right) \vert e_{2} \rangle \vert \mu_{0}^{1} \mu_{0}^{2} \rangle \nonumber \\ \nonumber
\\
+  \sqrt{\dfrac{p}{3}} \left( \alpha  \vert \phi^{+} \rangle  - \beta \vert \psi^{+} \rangle  \right) \vert e_{3} \rangle \vert \mu_{0}^{1} \mu_{0}^{2} \rangle
\label{decoherence1}
\end{align}
where $ p $ denotes the error probability.
Equation \eqref{decoherence1} describes the depolarizing dynamics suffered by the logical qubit, with the first term proportional to $ \sqrt{1-p}$ corresponding to no decoherence and the subsequent terms proportional to $ \sqrt{p/3} $ corresponding to errors on the logical qubit.
Note that at this stage, the corrective devices remain unaffected while the system undergoes errors and the environment \textit{learns} when an error has occurred. 
Tracing out the environment, the above evolution induces a dissipative map on the spin system increasing its entropy and causing decoherence of the original state. 

With the occurrence of errors the potential landscapes acting on $ b $ undergo a change forcing the action of the correctors upon the spin state of the nucleus. Purity of the logical qubit is restored at the expense of an increase in entropy for the correctors; after a correction event, \eqref{decoherence1} evolves to
\begin{align}
\vert \Psi \rangle \left(  \sqrt{1-p}  \vert e_{0} \rangle \vert \mu_{0}^{1} \mu_{0}^{2} \rangle +  \sqrt{\dfrac{p}{3}}  \vert e_{1} \rangle \vert \mu_{0}^{1} \mu_{1}^{2} \rangle \right. \nonumber \\ 
\left. - \sqrt{\dfrac{p}{3}}  \vert e_{2} \rangle \vert \mu_{1}^{1} \mu_{0}^{2} \rangle - \sqrt{\dfrac{p}{3}}  \vert e_{3} \rangle \vert  \mu_{1}^{1} \mu_{1}^{2} \rangle \right)
\end{align}
%It is interesting to note that the logical qubit state is recovered at the cost of entangling the loops with the environment and consequently increasing their entropy. 
where we can see that the original logical qubit state is restored and the environment gets correlated to the correctors' state.
After a single error correction, the correctors' states must be reset to the pure initial state $ \vert \mu_{0}^{1} \mu_{0}^{2} \rangle $. This is a non-unitary operation which requires energy expenditure, similar to erasing a quantum state \cite{Bennett2003, Bub2001} and can be implemented as a non-equilibrium stochastic process. 
This corresponds to a consumption of resources by the qubot analogous to the consumption of resources by biological molecular machines and living organisms. Irrespective of the physical implementation of the corrective sites, such consumption of resources is a mandatory part of the qubot operation in accordance to the laws of thermodynamics.
%This \textit{forgetness} operation is denoted $ \mathcal{F} $. 

\section{Implementation}

% Why Rydberg atoms? They have strong interactions. The VdW interaction for Rydberg states has an enhancement factor of n^4; for n = 60 that is 7 orders of magnitude higher than VdW interactions for the ground state. On the other hand, Rydberg states are unstable due to spontaneous decay, with lifetimes on the order of microseconds. By dressing ground states with a small fraction in probability amplitude of Rydberg states, one can achieve strong VdW interactions and long lifetimes. Interactions can prevail all the way to micron sized distances.

\textit{Potential engineering}. Spin-spin interactions of the form \eqref{interaction} suitable for implementing quantum robots could be engineered in a number of different atomic and molecular systems.
In this section a physical implementation using laser-dressed Rydberg atoms \cite{Glaetzle2015, Bijnen2015, Jau2015} is discussed. As will be shown, instances of the qubot described in the previous section can be realized for realistic experimental parameters, provided one chooses the correct logical basis elements and position of corrective sites. We will focus on a qubot that stabilizes an effective entangled spin state against a depolarizing environment similar to the one outlined in \cite{Guerreiro2020}. We shall refer to this device as an \textit{entanglement qubot}.

\begin{figure}[ht!] % [width=8.6 cm]
%    \centering
    \includegraphics[width =0.25\textwidth]{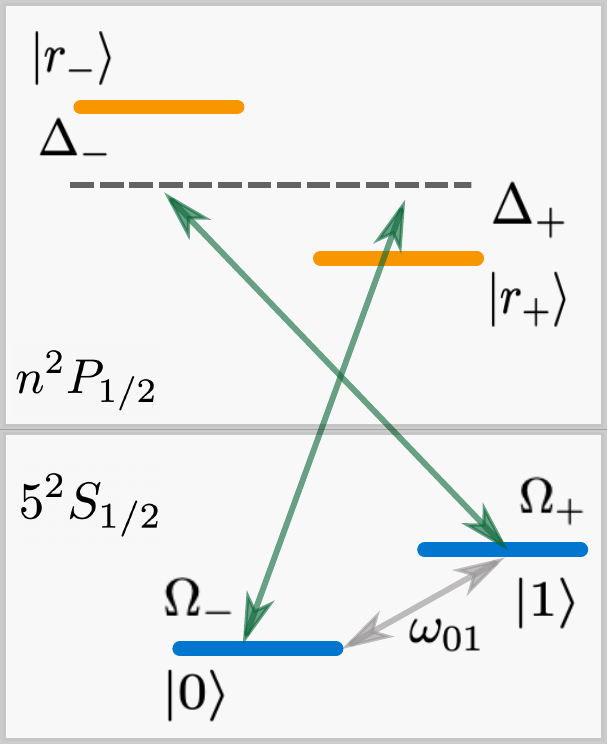}
    \caption[]{Level schematics for the entanglement qubot. }\label{levels}
    \label{color}
\end{figure}

\begin{figure*}[ht!] % [width=8.6 cm]
%    \centering
    \includegraphics[width =\textwidth]{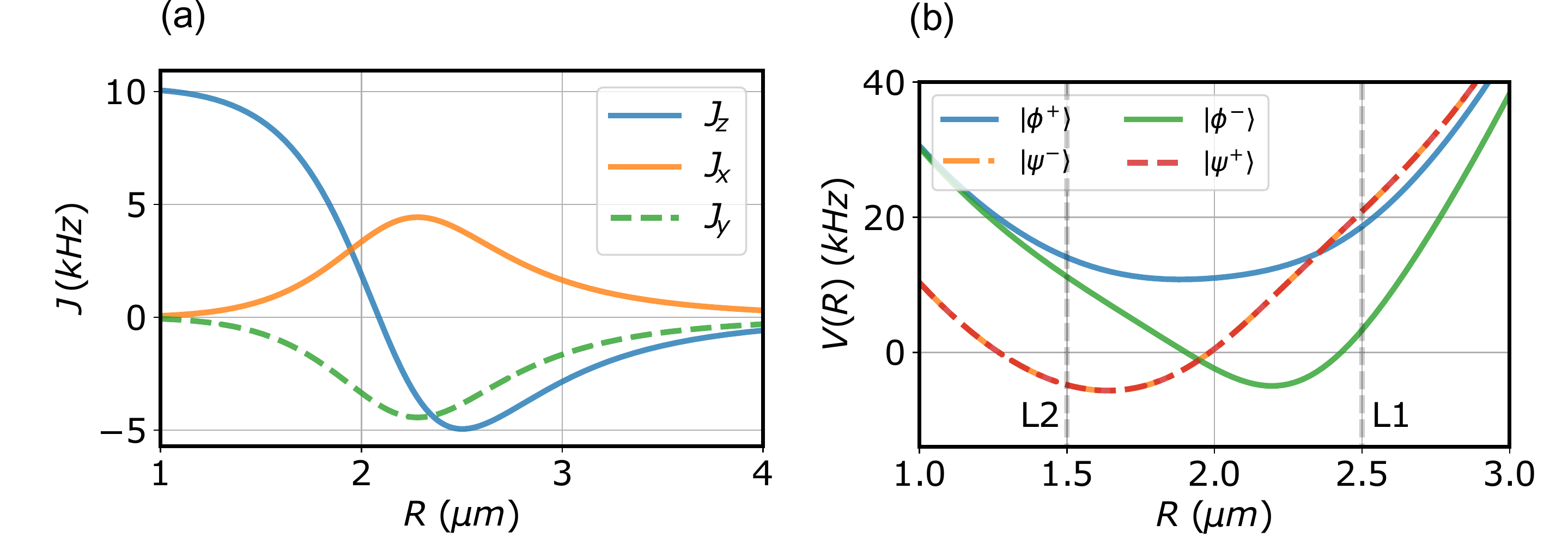}
    \caption[]{(a) Spin pattern, corresponding to the coefficients of Eq. \eqref{real_int} for the parameters $ n = 60 $, $ \Delta_{-} = - \Delta_{+} = 2\pi \times \SI{50}{MHz}  $ and $ \Omega_{-} = \Omega_{+} / 3 = \SI{2 \pi \times 3}{MHz} $. (b) Collective spin-dependent potential landscapes. Each trace corresponds to a Bell state of the qubot nucleus. Corrective devices $ L1 $ and $ L2 $ are positioned outside the potential minima, for example as the dashed vertical lines indicate. }\label{Rydberg_potentials}
    \label{color}
\end{figure*}

A pair of $ ^{87} $Rb atoms labelled $ a $ and $ b $ constitute the qubot nucleus. 
%The atoms, labelled $ a $ and $ b $, are trapped in a one-dimensional potential. 
Effective spin states are provided by hyperfine levels of $ b $, specifically
\begin{eqnarray}
\vert 0 \rangle &=& \vert 5^{2} S_{1/2}, F = 1, m_{F} = 1 \rangle \ , \\
\vert 1 \rangle &=& \vert 5^{2} S_{1/2}, F = 2, m_{F} = 2 \rangle \ ,
\end{eqnarray}
with energy difference $ \omega_{01} $.
The atom-atom interaction potential is induced by dressing the $ \vert 0 \rangle, \vert 1 \rangle $ states with two strongly interacting Rydberg Zeeman sublevels in the $n^{2} P_{1/2}$ manifold via Rabi oscillations with detunings $ \Delta_{\pm} $ and frequencies $ \Omega_{\pm} $ using $ \sigma^{\pm} $ polarized light. 
The interaction between Rydberg states arises from a van der Waals potential of the form $ C_{6} R^{-6} $, and a fixed orientation of the two particles is considered, with the atoms polarized perpendicular to the plane. 
Large detunings guarantee that only a small fraction of the Rydberg states is admixed to the $ \vert 0 \rangle, \vert 1 \rangle $ levels while maintaining a long lifetime. 
Following \cite{Glaetzle2015}, the Rydberg states are
\begin{eqnarray}
\vert r_{\pm} \rangle = \vert n^{2} P_{1/2}, m_{j} = \pm 1/2 \rangle \vert m_{I} = 3/2 \rangle \ ,
\end{eqnarray}
with an energy difference $ \Delta E_{r} $. Detunings are chosen such that the energy conservation condition $ \Delta E_{r} = (\Delta_{+} - \Delta_{-})  $ is satisfied. 
A level diagram is shown in Figure \ref{levels}. 
The atoms are trapped in one dimensional potentials, insensitive to their internal states. State-independent trapping of Rydberg dressed atoms can be achieved in so-called magic \cite{Ye2008, Zhang2011} and magnetic traps \cite{Boetes2018}. While atom $ a $ is fixed at the origin, $ b $ is able to move under the influence of a force resulting from the combination of an external tweezer and the atom-atom interaction potential.

As in quantum chemistry \cite{Carr2009, Krems}, the time scale associated to electronic dynamics is much shorter than the time scale of nuclei motion. An effective spin dependent Born-Oppenheimer potential can therefore be derived at fixed atomic separations $ R $.
In the limit of large detunings $ \Omega_{\pm} \ll \Delta_{\pm}  $ and for  $ \Delta_{+}/ \Delta_{-} < 0, \Delta_{+} + \Delta_{-}< 0 $, adiabatic elimination \cite{Paulisch2014} can be used in the rotating frame to obtain an effective interaction acting on the subspace generated by the $ \vert 0 \rangle, \vert 1 \rangle $ states to fourth order in $ \Omega_{\pm} / \Delta_{\pm} $, 
%derive an effective spin-dependent Born-Oppenheimer 
\begin{align}
V_{I}(R) = J_{z} Z_{a} Z_{b} + J_{x} X_{a} X_{b} + J_{y} Y_{a} Y_{b} +J_{\parallel} \left( Z_{a} + Z_{b} \right) \ ,
\label{real_int}
\end{align}
where $  J_{\alpha}(R) $ ($ \alpha = x,y,z $) are radial steplike coefficients depending on the Rabi frequencies $ \Omega_{\pm} $, detunings $ \Delta_{\pm} $ and van der Waals $ C_{6} $ coefficients for the $ n^{2} P_{1/2}$ manifold.  $ J_{\parallel} $ is an effective magnetic field, which we assume can be cancelled by an additional weak non-homogeneous field on the order of \SI{2}{G}. See Appendices A and B for explicit definitions, formulas and details on the potential and effective magnetic field, respectively.

A plot of the $ J_{\alpha} $ spin pattern for $ n = 60 $, detunings $ \Delta_{-} = - \Delta_{+} = 2\pi \times \SI{50}{MHz}  $ and Rabi frequencies $ \Omega_{-} = \Omega_{+} / 3 = \SI{2 \pi \times 3}{MHz} $ can be seen in Figure \ref{Rydberg_potentials}(a). Note these are in the same parameter region as used for realizing the quantum spin ice Hamiltonian on a kagome lattice in \cite{Glaetzle2015, Glaetzle2014}. The parameters defining a qubot potential are not unique, allowing some freedom in the construction; for an example of a different set of numbers and the resulting spin pattern see the Appendix C. 

From the spin pattern coefficients together with Eqs.\eqref{V_1}-\eqref{V_4} and a trap potential $ V_{t}(R) $ we can derive the collective spin-dependent potentials acting on particle $ b $. 
%The effective magnetic field provided by the last term in \eqref{real_int}  is on average $ \langle J_{\parallel} \rangle \approx \SI{1780}{kHz} $. Considering the Land\'e factor $ \vert g_{F} \vert \approx \SI{0.70}{MHz / G} $ for the $ 5^{2} S_{1/2} $ states \cite{Bize1999} this effective magnetic field can be cancelled by an additional weak non-homogeneous field of order of magnitude $ \vert B_{c} \vert \approx \SI{3}{G} $.
%We will therefore neglect the last term in Eq. \eqref{real_int}. 
Consider a trap potential provided by two neighboring optical tweezers,
\begin{eqnarray}
V_{t}(R) = V_{0} \left[ \left(  R - \delta_{1} \right)^{2} + \left(  R - \delta_{2} \right)^{2}    \right]
\end{eqnarray}
where $ V_{0} = \SI{15}{kHz / \mu m^{2}} $, $ \delta_{1} = \SI{1.6}{\mu m} $ and $ \delta_{2} = \SI{2.0}{\mu m} $. 
The resulting spin-dependent potential landscapes $ V(R) $ can be seen in Figure \ref{Rydberg_potentials}(b), where each trace corresponds to a different Bell state of the two atoms. Note equilibrium positions are separated by approximately $ \SI{0.3}{\mu m} $. Trap frequencies are approximately $ \omega_{t} / 2 \pi \approx  \SI{1}{kHz} $.
Possible positions for the corrective sites $ L1 $ and $ L2 $, corresponding to the transformations \eqref{loop_eqs}, are represented by dashed vertical lines. 
Note the potential landscapes for the Bell states $ \vert \psi^{-} \rangle $ and $ \vert \psi^{+} \rangle $ overlap. This implies that one cannot choose either $ \vert \psi^{-} \rangle $ or $ \vert \psi^{+} \rangle $ as protected states, as in this case, phase errors could not be corrected.
%cannot imparts on the capacity of the qubot of protecting these states. 
The protected logical state is chosen to be $ \vert \phi^{+} \rangle $.

\textit{Corretive sites}. Correctors $ L1 $ and $ L2 $ were previously considered to be qubits acting as an entropy sink for maintaining the purity of the protected logical qubit state carried by the nucleus. The interaction between superconducting quantum electronics and atomic \cite{Kielpinski2012}, molecular \cite{Andre2006} and mesoscopic particles \cite{Martinetz2020} has been extensively studied in the context of hybrid quantum systems and the coupling between NV centers and superconductors has been observed \cite{Kubo2011}. A number of different implementations involving superconducting qubit systems is therefore expected.

Beyond qubits, one may consider additional atoms as candidates for implementing corrective devices. Controlled atomic collisions \cite{Jaksch1999} would provide the mechanism for position-dependent unitary operations. 
One could envision a lattice with arrays of \textit{data} particles interpolated with \textit{corrective} particles, analogous to the surface code \cite{Fowler2012}; occurrence of errors would alter the interaction between data particles, enabling or inhibiting motion and tunneling - and consequently interactions - with neighboring corrective sites. 
It would be as a surface code \textit{in motion}, where errors induce controlled motion leading to correction feedbacks. 
It is important to stress that in the course of the qubot action, entropy of the corrective atoms would increase and a dissipative map for restarting the correctors in their original state would have to be continuously enforced, for example through an amplitude damping channel \cite{Guerreiro2020}. 

\begin{figure}[ht!] % [width=8.6 cm]
%    \centering
    \includegraphics[width =0.4\textwidth]{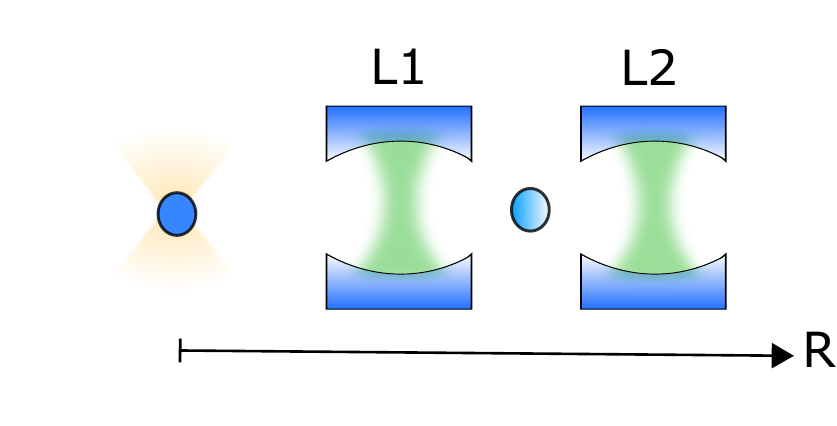}
    \caption[]{Corrective sites as Rabi oscillations. }\label{setupB}
    \label{color}
\end{figure}

Corrective devices could also be implemented using Rabi oscillations between the $ \vert 0 \rangle, \vert 1 \rangle $ levels. By carefully tuning the Rabi frequency of the transition and the profile of the spin-dependent potentials in Figure \ref{Rydberg_potentials}(b) it is in principle possible to engineer the transit time of atom $ b $ through $ L1 $ and $ L2 $ such that $ Z_{b} $ and $ X_{b} $ operations are applied, analogous to the transit time stimulated decay in ammonia masers \cite{Feynman_lect} and Ramsey interferometry in atomic fountain clocks \cite{Wynands2005}. In this implementation - probably the most practical from an experimental point-of-view - the electromagnetic field assumes the role of entropy sink since conditional $ X $ and $ Z $ operations on the atom would introduce uncertainties in the intensity and phase of the field, respectively. A schematics of this implementation is shown in Figure \ref{setupB}.

\textit{Operation, cooling and lifetime}. Operation of the qubot proceeds as described in the previous section: occurrence of an error induces a change in the potential landscape seen by atom $ b $ thus forcing it into one of the corrective sites $ L1 $ or $ L2 $.  
Note that errors can occur due to external environmental influence or intrinsically due to thermal and quantum fluctuations of the atomic motion. 
Consider atom $ b $ in a thermal state. For temperatures on the order of \SI{10}{nK}, reachable for atomic ensembles \cite{Weld2009}, the occupation number of atomic motion is $ \bar{n} \approx 0.1 $ pointing out that the atom is effectively in the trap ground state. 
%It is therefore interesting to compare the thermal rms to the quantum fluctuations due to the approximatelly harmonic potential ground state. 
Zero point motion of the atom is approximately $ R_{\mathrm{zpm}} \simeq \sqrt{\hbar/2m\omega_{t}} \approx \SI{0.23}{\mu m} $, indicating that at \SI{10}{nK} quantum fluctuations can cause the atom to reach the corrective sites even when no environmental error took place, inducing change in the qubot state. Hence, intrinsic fluctuation errors are expected to constitute a portion of total errors. In the next section, a model of the qubot operation taking into account intrinsic and external errors will be discussed.

Errors can be effectively corrected provided the qubot nucleus undergoes constant cooling of its motional degrees of freedom to dissipate the kinetic energy gained by mechanical forces due to potential changes.
Such cooling mechanism needs to preserve the quantum information stored in the nucleus, so it must be insensitive to the quantum state stored in the spins. State-insensitive cooling of neutral atoms can be achieved via superfluid immersion \cite{Daley2004}, cavity cooling \cite{Griessner2004} or sympathetic cooling through spin-independent Rydberg interactions with neighboring atoms \cite{Belyansky2019}.  

What is the order of magnitude of the expected lifetime for the protected entangled state? The $ 60P_{1/2} $ Rydberg state has a lifetime on the order of $ \tau_{r} \approx \SI{133}{\mu s} $ \cite{Beterov2009}. This implies a bare lifetime for the effective spin state of $ \tau_{s} \approx (2 \Delta_{-}/\Omega_{-})^{2} \tau_{r} \approx \SI{9}{ms} $ \cite{Glaetzle2015}, corresponding to a spin decoherence rate $ \Gamma \approx \SI{111}{Hz} $. A decay process to the ground state $ \vert 0 \rangle $ is defined by the following transformations,
\begin{eqnarray}
\vert 0 \rangle \vert e_{0} \rangle &\rightarrow & \vert 0 \rangle \vert e_{0} \rangle \\
\vert 1 \rangle \vert e_{0} \rangle &\rightarrow & \sqrt{1 - \tau_{s}^{-1} dt}  \vert 1 \rangle \vert e_{0} \rangle + \sqrt{\tau_{s}^{-1} dt} \vert 0 \rangle \vert e_{1} \rangle
\end{eqnarray}
where the first ket corresponds to the spin of the particle while the second ket represents the environment state. 
Action of this quantum channel upon the elements of the Bell basis can be written in terms of strings of Pauli errors \cite{Preskill}. 
%Note that the decay process corresponds to an application of the $ X $ operator on the qubit state.
%The spin singlet state of the joint $ ab $ system transforms under these operations as
%\begin{eqnarray}
%\vert \psi^{-} \rangle \vert e_{0} \rangle & \rightarrow & \sqrt{1 - \tau_{s}^{-1} dt}  \vert \psi^{-} \rangle \vert e_{0} \rangle  +  \nonumber \\ 
% &+& \sqrt{\tau_{s}^{-1} dt}  \left(  \dfrac{ i Y_{b}  + X_{b}}{2}  \right) \vert \psi^{-} \rangle \vert e_{1} \rangle
%\end{eqnarray}
%showing that decay to the ground state consists of a superposition of $ X $ and $ Y $ errors. 
It is thus expected that the qubot is able to extend the lifetime of Rydberg dressed entangled states.
%; to quantitatively evaluate the protection we proceed to simulate the system dynamics.

\section{Dynamics Simulation}

% Random motion with probability distribution as predicted by the Schrodinger equation; it's a treatment that describes a quantum system being continually observed, yielding `quantum trajectories'. 

Exploration of the qubot requires simulation of its error-correction dynamics.
Any such simulation must take into account the effects of quantum fluctuations of atomic motion, as these fluctuations are in themselves a source of errors that can disturb the protected Bell state. 
A first principles description of the spin and motion degrees of freedom is intricate as the spin state is subject to transformations conditional on the motion state, which in itself is conditioned on the spin through the spin-dependent potential. 
As Wheeler would say \cite{Wheeler}: \textit{spin tells matter how to move, matter tells spin how to turn}. 

To capture the essential features of the qubot we propose an open quantum system model in which the motion and spin degrees of freedom follow a set of discrete-time coupled stochastic Schrodinger equations.
%These stochastic equations correspond to the unraveling of a master equation, and mean behavior of the system can be calculated by averaging over many realizations of the stochastic process. 
%Physically, it makes sense to talk about a single realization of the qubot. Hence the coupled master equations serve to motivate a stochastic wavefunction approach. A quantum Monte-Carlo simulation is then used to predict the mean behavior of the system.
Each realization of the evolution is described in terms of sequences of quantum state pairs, denoted $ \vert \psi \rangle $ for the spin and $ \vert \phi \rangle $ for the motion degree of freedom. 
Averaging over many realizations of the stochastic process results in the mean behavior of the system.

The spin and motion degrees of freedom act as environments for each other. This idea can be used to motivate the model as follows.
%Say we start with a spin-motion state given by $ \vert \psi \rangle \vert \phi \rangle $.
For simplicity, discretize (1D) space into a set of points $ R_{k} $. The position state reads
\begin{eqnarray}
\vert \phi \rangle = \sum_{k} \phi(R_{k}) \vert R_{k} \rangle
\end{eqnarray}
where $ \vert \phi(R_{k}) \vert^{2} $ gives the probability of finding the particle at position $ R_{k} $.
The initial state evolves in a small time increment $ \delta t $ according to
\begin{align}
\vert \psi \rangle \vert \phi \rangle \xrightarrow{\delta t}  \sum_{i} \phi(R_{i}) ( T(R_{i}) \vert \psi \rangle ) (W(\vert \psi \rangle) \vert R_{i} \rangle )  \nonumber \\
= \vert \Psi(t + \delta t) \rangle 
\end{align}
where $ T(R_{i}) $ is the identity operator unless $ R_{i} = R_{L1} $ or $ R_{i} = R_{L2} $, for which
\begin{eqnarray}
T(R_{L1}) = Z_{b} \\
T(R_{L2}) = X_{b}
\end{eqnarray}
The operator $ W(\vert \psi \rangle) $ contains information on the spin-dependent potential and is responsible for the evolution of the motion state. 
Expanding $ \vert \Psi(t + dt) \rangle $,
\begin{eqnarray}
\vert \Psi(t + \delta t) \rangle &=& \sum_{i \neq L1, L2} \phi(R_{i})  \vert \psi \rangle  (W(\vert \psi \rangle) \vert R_{i} \rangle ) \nonumber  \\
&+& \phi(R_{L1}) ( Z_{b} \vert \psi \rangle ) (W(\vert \psi \rangle) \vert R_{L1} \rangle ) \nonumber \\
&+& \phi(R_{L2}) ( X_{b} \vert \psi \rangle ) (W(\vert \psi \rangle) \vert R_{L2} \rangle )
\label{unitary_spin_motion}
\end{eqnarray}
%This expression motivates our model of coupled discrete time stochastic Schrodinger equations. 
%When assuming the spin and motion states follow coupled stochastic equations we are ignoring phase information and correlations of the global state. This is justified when the spin state is constantly measured in the Bell basis by an external environment.
Assuming the spin state is continuously monitored in the Bell basis, the above state continuously collapses to a random separable state allowing the phase information and correlations of the global state to be ignored. Note that under this \textit{monitoring} assumption one can describe the dynamics of the system within a simpler scenario and yet verify the error correction capability of the proposed qubot. Moreover, monitoring of the joint spin state in the Bell basis can be achieved by continuous measurement of the force acting on particle $ a $, since the interaction between the particles is given by their joint spin state.
The motion state then acts as an environment for the spin, inducing \textit{corrective} jump operators,
\begin{eqnarray}
L_{1} = \sqrt{\gamma_{L1}}   Z_{b} \\
L_{2} = \sqrt{\gamma_{L2}}   X_{b}
\end{eqnarray} 
where we define \textit{correction rates} as
\begin{eqnarray}
\gamma_{L1} dt &=& \vert \phi(R_{L1})\vert^{2} \\
\gamma_{L2} dt &=& \vert \phi(R_{L2}) \vert ^{2} 
\end{eqnarray}
Note that the probability of a given corrective jump occuring is also the probability of finding the particle in the corresponding corrective site.
In addition to corrective jumps the spin state is also under the effect of a depolarizing channel due to an external decoherence environment, defined in terms of the collapse operators
\begin{eqnarray}
L_{3} = \sqrt{\dfrac{\Gamma}{3}} X_{b} \ , \  L_{4} = \sqrt{\dfrac{\Gamma}{3}} Y_{b} \ , L_{5} = \sqrt{\dfrac{\Gamma}{3}} Z_{b} \ , \ 
\end{eqnarray} 
where $ \Gamma $ is the decoherence rate. 

Conversely spin acts as an environment to the motion state. If no spin corrective jump occurs the motion state is left almost unperturbed, according to \eqref{unitary_spin_motion}, and evolves through the unitary predicted by the spin state $ \vert \psi \rangle $ plus the effects of a damping collapse operator provided by an additional spin-insensitive cooling environment with damping rate $ \kappa $ acting as a drain of kinetic energy, as discussed previously. 
On the other hand, if a corrective jump $ L_{1} $ or $ L_{2} $ happens the motion state collapses to $  \vert R_{L1} \rangle $ or $ \vert R_{L2} \rangle $, respectively.  The collapsed state subsequently evolves according to the unitary predicted by the spin state $ \vert \psi \rangle $ plus the additional damping collapse operator. 
When spin jumps happen, the motion Hamiltonian must be updated accordingly for the next time iteration. 

This evolution can be implemented via a \textit{coupled} Monte-Carlo method. First, define the Motion Monte-Carlo procedure (MMC) for a damped harmonic oscillator as following:
\begin{itemize}

\item[\textbf{(1)}] Define motion state $ \vert \phi \rangle $ and Hamiltonian $ H $;

\item[\textbf{(2)}] Compute $ \delta v = \kappa \delta t \langle \phi \vert a^{\dagger} a \vert \phi \rangle $;

\item[\textbf{(3)}] Choose uniformly distributed random number $ q \in  [0,1] $;

\item[\textbf{(4)}] If $ q < \delta v $, update $\vert \phi \rangle \leftarrow a\vert \phi \rangle / \sqrt{ \delta v / \delta t} $;

\item[\textbf{(5)}] If $ q > \delta v $, update $ \vert \phi \rangle \leftarrow e^{-i \hat{H} \delta t} \vert \phi \rangle / \sqrt{1 - \delta v} $, where $ \hat{H} = H - \frac{i}{2} a^{\dagger} a $;

\end{itemize}
We denote by \textbf{MMC}$(\vert \phi \rangle, H, \delta t) $ the output of the above procedure for input state $ \vert \phi \rangle $, Hamiltonian $ H $, over a time step $ \delta t $. This output consists of the updated motion state after one time step.

The following algorithm, dubbed Spin-Motion Monte Carlo (\textbf{SMMC}), summarizes one time iteration of the qubot dynamics:

\begin{itemize}
\item[\textbf{(1)}] Define (update) motion and spin states $ \vert \phi \rangle $ and $  \vert \psi \rangle $ and motion Hamiltonian $ H = H(\vert \psi \rangle) $;

\item[\textbf{(2)}] Define correction rates $ \gamma_{L1} \delta t = \vert \langle R_{L1} \vert \phi \rangle \vert^{2} , \gamma_{L2} \delta t = \vert \langle R_{L2} \vert \phi\rangle \vert^{2} $, where $ \delta t $ is the discrete time increment;

\item[\textbf{(3)}]  Compute $ \delta p_{k} = \delta t \langle \psi \vert L_{k}^{\dagger} L_{k} \vert \psi \rangle $ and $ \delta p = \sum_{k} \delta p_{k} $;

\item[\textbf{(4)}] Choose uniformly distributed random number $ r \in  [0,1] $;

\item[\textbf{(5)}] If $ r < \delta p $, update $\vert \psi \rangle \leftarrow L_{k} \vert \psi \rangle / \sqrt{dp_{k}/ \delta t} $ with probability $ \delta p_{k} / \delta p $;

\textbf{(5.1)} If jumps $ L_{k} $ with $ k = 1 $ or $ 2 $ occurred, update $ \vert \phi \rangle \leftarrow \vert R_{L_{k}} \rangle $ and run  \textbf{MMC}$(\vert R_{L_{k}} \rangle, H,\delta t)$. After MMC update the motion state and the motion Hamiltonian to $ H = H(L_{k}\vert \psi \rangle) $;

\textbf{(5.2)}  If jumps $ L_{k} $ with $ k = 3, 4 $ or $ 5 $ occurred, run  \textbf{MMC}$(\vert \phi \rangle, H,\delta t)$. After MMC update the motion state and the motion Hamiltonian to $ H = H(L_{k}\vert \psi \rangle) $;

\item[\textbf{(6)}] If $ r > \delta p $, update $ \vert \psi \rangle \leftarrow e^{-i H_{s} \delta t} \vert \psi \rangle / \sqrt{1 - \delta p} $, where $ H_{s} = -i \sum_{k} L_{k}^{\dagger} L_{k} $;

\textbf{(6.1)} Run  \textbf{MMC}$(\vert \phi \rangle, H,\delta t)$. After MMC update the motion state and the motion Hamiltonian to $ H = H(  \vert \psi\rangle  )  $;

\item[\textbf{(7)}] Go to \textbf{(1)} for next iteration.

\end{itemize}
A time series of quantum states $ \lbrace \vert \psi(t) \rangle, \vert \phi(t) \rangle \rbrace $ is called a quantum trajectory of the system, and can be obtained by iterating \textbf{SMMC}. Mean behavior of the qubot can be obtained by averaging quantities of interest over many quantum trajectories. For example, we can define the \textit{overlap} between the qubot spin state and the protected Bell state as $ F = \mathbb{E} \left[ \vert \langle \psi(t) \vert \phi^{+} \rangle \vert^{2} \right]  $, where $\mathbb{E} \left[ ... \right] $ denotes the ensemble average over all quantum trajectories. The quantity $ F $ then measures how close the qubot spin state is on average to the protected state and hence quantifies how well the qubot functions. 

To simplify the dynamics simulation, spin-dependent potentials are taken to be harmonic traps of equal resonance frequency. This removes any issues due to anharmonicity in the potentials and allows for the definition of fixed phonon creation and annihilation operators. 
The potentials shown in Figure \ref{Rydberg_potentials}(b) are approximated as
\begin{eqnarray}
V(\vert \psi \rangle ,R) = \dfrac{m  \omega_{t}^{2}}{2} \left[ R - R_{0}(\vert \psi \rangle) \right]^{2}
\end{eqnarray}
where $ \omega_{t} / 2 \pi = \SI{1}{kHz} $ and the trap position $ R_{0}(\vert \psi \rangle) $ is given by

\begin{figure}[ht!] % [width=8.6 cm]
%    \centering
    \includegraphics[width =0.46\textwidth]{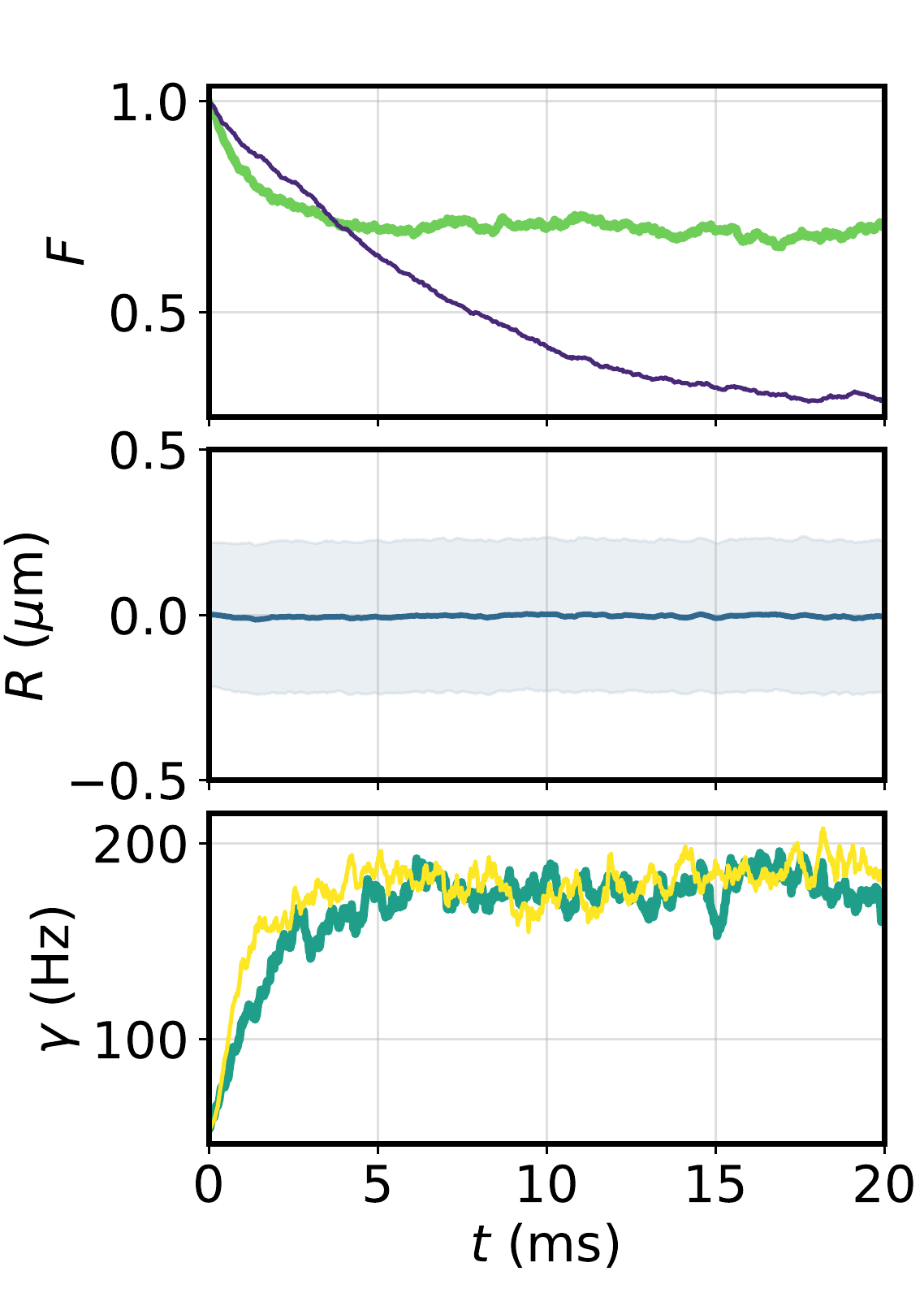}
    \caption[]{Coupled spin-motion Monte-Carlo simulation of the qubot, $ 10^{3} $ quantum trajectories. Top: average fidelity to the $ \vert \phi^{+} \rangle $ Bell state as a function of time for the qubot plus a depolarizing channel (thick green line) compared to the action of a depolarizing channel alone (thin purple line). Middle: average position of the atom with corresponding quantum uncertainty (light blue shade). Bottom: average correction rates $ \gamma_{L1} $ (light yellow line) and $ \gamma_{L2} $ (thick green line). The parameters used in the plot are: decoherence rate $ \Gamma = \SI{100}{Hz} $, trap frequency $ \omega_{t} = \SI{1}{kHz} $, damping rate $ \kappa = \SI{0.1}{ms} \times \omega_{t}^{2}  $, initial wavepacket uncertainty $ \Delta R = \SI{0.22}{\mu m} $, $ R_{L2} = - R_{L1} = \SI{0.63}{\mu m} $.}\label{simulation1}
    \label{color}
\end{figure}

\begin{eqnarray}
R_{0}(\vert \psi \rangle) = \left\{ \begin{array}{ll}
      R_{01}, & \mathrm{if} \ \vert \phi^{+} \rangle \\
      R_{10}, & \mathrm{if} \ \vert \phi^{-} \rangle  \\
      R_{00}, & \mathrm{if} \ \vert \psi^{\pm} \rangle  \\
\end{array} 
\right.
\end{eqnarray}
The positions $ R_{\alpha \beta} $ are dependent on the details of the experimental implementation. 
Inspired by Figure \ref{Rydberg_potentials}(b) we consider $ R_{01} = \SI{1.90}{\mu m} $, $ R_{10} = \SI{2.20}{\mu m} $ and $ R_{00} =\SI{1.64}{\mu m}  $.
%We wish to define master equations and unravel them into stochastic Schrodinger equations for numerical simulation of quantum trajectories. As will later become clear, this stochastic wavefunction approach only requires definitions of $ R_{0}(\rho_{s}) $ for the four Bell states. 
Since the Hamiltonian always appears inside a commutator, constant terms can be neglected without affecting the dynamics. Defining the origin of our reference frame at the minimum of the potential $ V(\vert \phi^{+} \rangle) $ and neglecting constant shifts, the Hamiltonian reads
\begin{eqnarray}
H(\vert \psi \rangle) = \omega_{t} a^{\dagger} a -  m  \omega_{t}^{2} \Delta R_{0}( \vert \psi \rangle ) R_{\mathrm{zpm}} \left( a^{\dagger} + a  \right)
\label{Hamiltonian_sim}
\end{eqnarray}
with $ a^{\dagger}, a $ the creation and annihilation operators for the $ \vert \phi^{+} \rangle $ potential, given by,
\begin{eqnarray}
a = \sqrt{\dfrac{m\omega_{t}}{2}} \left( R + \dfrac{i}{m\omega_{t}} P   \right) \\
a^{\dagger} = \sqrt{\dfrac{m\omega_{t}}{2}} \left( R - \dfrac{i}{m\omega_{t}} P   \right)
\end{eqnarray}
with $ R, P $ the atom position and momentum operators of particle $ b $, respectively, $ R_{\mathrm{zpm}} $ the corresponding zero-point motion and $ \Delta R_{0}(\vert \psi \rangle) = R_{0}(\vert \psi \rangle) - R_{0}(\vert \phi^{+} \rangle  ) $. 
The effect of a change in the spin state can be interpreted as the appearance of an additional force acting on particle $ b $.

Figure \ref{simulation1} shows the result of iterating \textbf{SMMC} averaged over $ 10^{3} $ quantum trajectories, implemented using QuTiP \cite{qutip}, for the initial Bell-position state $ \vert \phi^{+} \rangle \vert \chi \rangle $, where $ \vert \chi \rangle $ is a Gaussian wavepacket in position with uncertainty $ \Delta R $. 
See the Figure caption for details on the parameters used in the simulation.
The top graph shows the mean overlap $ F = \mathbb{E} \left[ \vert \langle \psi(t) \vert \phi^{+} \rangle \vert^{2} \right]  $ as a function of time for the qubot (thick green line) compared to the depolarizing channel alone (thin purple line). We can see that initially the qubot overlap drops faster than the free spins, but it stabilizes at about $ 70\% $, while free decohering spins decrease significantly below. 
%The fast drop in overlap is due to intrinsic qubot decoherence caused by quantum fluctuations of the atomic motion and can be minimized by increasing $ \vert R_{L1} \vert $ and $ \vert R_{L2} \vert $ at the expense of reducing correction rates.
The middle plot shows the atom position and its quantum uncertainty as a function of time: action of the qubot stabilizes the location of the atom. Note that motion of the atom towards one corrective site is expected to increase correction rates of that site and decrease correction rates of the other. This behavior can be seen in the bottom graph, where rates are shown as a function of time. As expected, $ \gamma_{L1} $ (light yellow line) displays significant anti-correlation with $ \gamma_{L2} $ (thick green line). 

The effect of finite temperature can be evaluated by adapting \textbf{SMMC} to include motion collapse operators $ \sqrt{\kappa(\bar{n} + 1)} a  $ and $ \sqrt{\kappa \bar{n}} a^{\dagger} $ representing contact with a thermal bath of phonons at temperature $ T $ with coupling $ \kappa $ and thermal occupation number $ \bar{n} $, where $ \bar{n} = 1 / (e^{\hbar \omega_{t} / k_{B} T} - 1) $. When in contact with a thermal bath, the particle initially in the ground state evolves to a thermal state with mean number of phonons $ \bar{n} $, increasing the position spread and consequently the intrinsic qubot error rate. The spin overlap is thus expected to decrease with temperature.
\begin{figure}[ht!] % [width=8.6 cm]
%    \centering
    \includegraphics[width =0.5\textwidth]{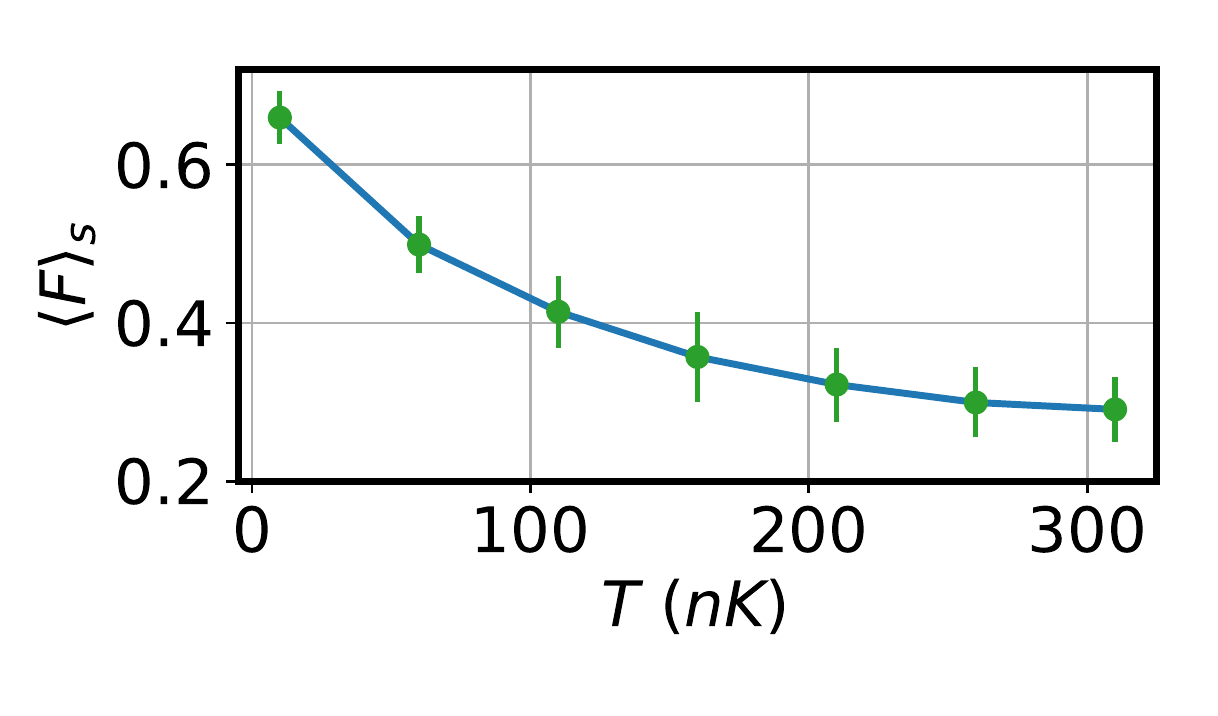}
    \caption[]{Effect of contact with a thermal bath at temperature $ T $ upon the steady state time-averaged overlap $ \langle F \rangle_{s} $. Time average is considered starting at \SI{10}{ms}, when the overlap has already achieved its steady value. Error bars correspond to one standard deviation. Each point is evaluated from $ 10^{2} $ quantum trajectories. Coupling to the heat bath is $ \kappa = \SI{0.1}{ms} \times \omega_{t}^{2}  $ and all remaining parameters are the same as in Figure \ref{simulation1}.}\label{Temperature}
    \label{color}
\end{figure}
The time-averaged steady state overlap $ \langle F \rangle_{s} $ as a function of temperature is plotted in Figure \ref{Temperature}. Each point is the result of time-averaging $ 10^{2} $ quantum trajectories with error bars corresponding to one standard deviation. As expected, the effect of contact with a heat bath is to decrease the overlap.

\begin{figure}[ht!] % [width=8.6 cm]
%    \centering
    \includegraphics[width =0.45\textwidth]{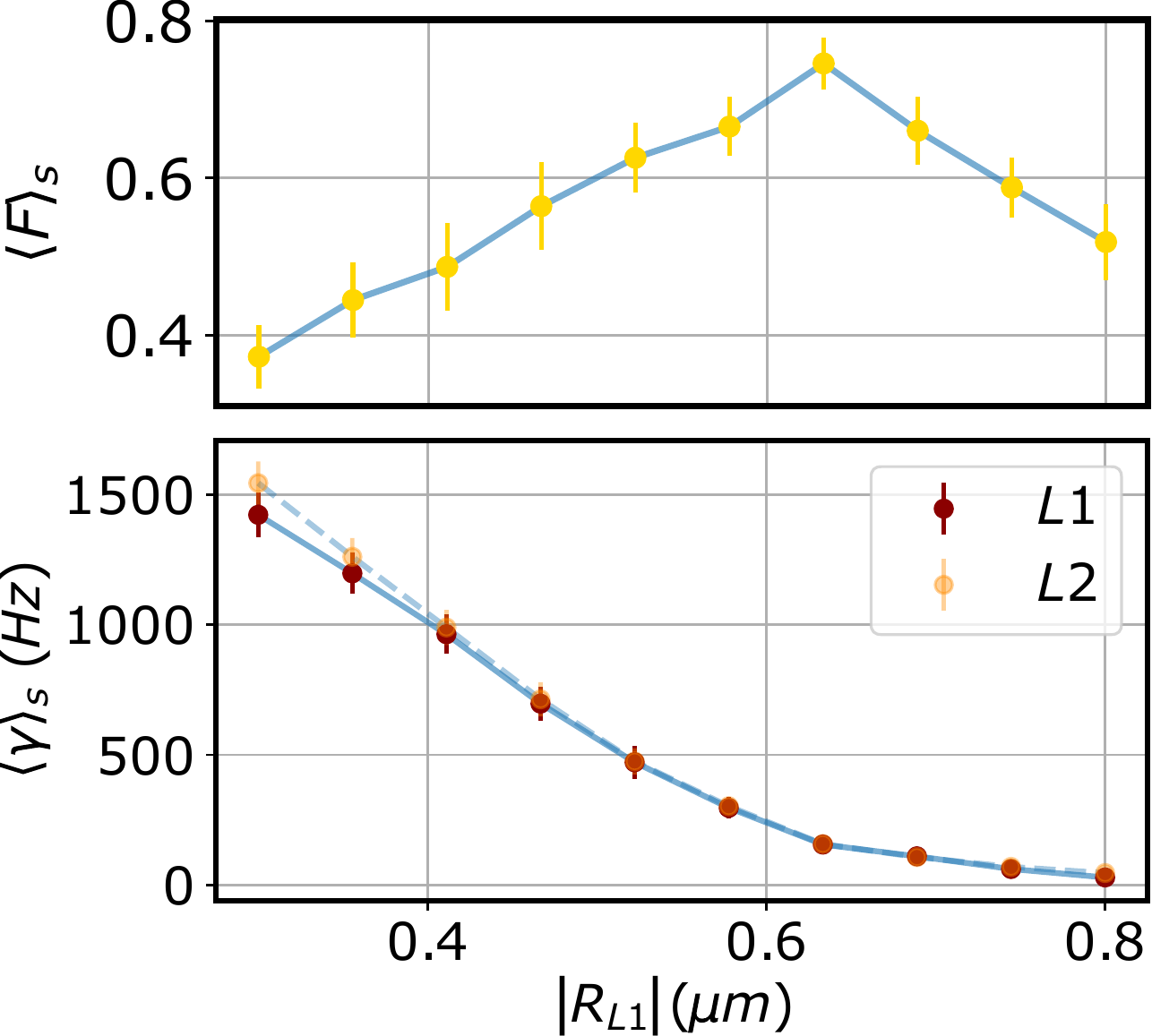}
    \caption[]{Influence of corrector positioning. Top: steady state overlap. Bottom: mean correction rates. Averages are considered from $\SI{10}{ms}$ onward, when the device is well settled in the steady state. Error bars correspond to one standard deviation. Each point is evaluated from $ 10^{2} $ quantum trajectories. All remaining parameters are the same as in Figure \ref{simulation1}.}\label{optimal}
    \label{color}
\end{figure}

Quantum fluctuations of the atomic motion can induce \textit{internal} errors if the atom interacts with the correctors when no external (decoherence) error has taken place. To quantify that effect, the steady state overlap $ \langle F \rangle_{s} $ and correction rates $ \langle \gamma \rangle_{s} $ are numerically calculated for different values of the $ L1 $ position $ \vert R_{L1} \vert $, shown in Figure \ref{optimal}; $ R_{L1}  = - R_{L2}  $ is assumed. Note that if the correctors are too close to the equilibrium position of $ \vert \phi^{+} \rangle ($ $\vert R_{L1} \vert < \SI{0.40}{\mu m}$), the steady state overlap $ \langle F \rangle_{s} $ falls below 50\%, while the mean rate for `correction' events are on the order of \SI{1}{kHz}, due to the atom fluctuating towards $ L1 $ or $ L2 $ even in the absence of an error. As $\vert R_{L1} \vert $ is increased, the steady state overlap increases, reaching a maximum value $  \langle F \rangle_{s} \approx 0.7 $ for $ \vert R_{L1} \vert \approx 0.63 $, and then decreases again as the correctors are placed further apart from the atom. 
The mean correction rates can be seen to decrease as the position $\vert R_{L1} \vert $ is further increased, which is intuitive since larger distances imply longer correction times.
The optimal operation point $ \vert R_{L1} \vert \approx 0.63 $ is such that the mean correction rates $ \langle \gamma \rangle $ are of the same order of the decoherence rate $ \Gamma = \SI{100}{Hz}$.
See Appendix D for more details.

%The fast drop in overlap is due to intrinsic qubot decoherence caused by quantum fluctuations of the atomic motion and can in principle be minimized by increasing $ \vert R_{L1} \vert $ and $ \vert R_{L2} \vert $ at the expense of reducing correction rates.

\section{Discussion}

Throughout this work we discussed quantum robots, devices as the one conceptualized in \cite{Guerreiro2020}, capable of harnessing interactions between its constituent parts and the surrounding environment to achieve targeted tasks such as state protection against decoherence. We have introduced for the first time a model of a qubot capable of partially protecting an arbitrary logical qubit state against general single physical qubit errors. The first physical implementation of an instance of such device, capable of protecting a Bell state against the detrimental action of a depolarizing environment has been described, as well as Monte-Carlo simulations of the qubot dynamics and the inclusion of effects due to contact of the device with a thermal bath.

%Throughout this work we discussed quantum robots, devices capable of harnessing interactions between its constituent parts and the surrounding environment to achieve targeted tasks such as state protection against decoherence.
%A conceptual model capable of partially protecting an arbitrary logical qubit state against single qubit physical errors was discussed. 
%Moreover, a specific implementation of a device inspired by the conceptual model was proposed. 
%The implementation is constructed to protect a Bell state state against the detrimental action of a dephasing environment.
%Such devices, recently conceptualized in \cite{Guerreiro2020}, have been extended to the realm of arbitrary logical qubit states subject to general physical qubit errors for the first time.

%\hl{The present work provides a concrete example of qubots, recently conceptualized in} \cite{Guerreiro2020}. \hl{It is important to note that the current proposal extends, for the first time, qubots to the realm of arbitrary logical qubit states subject to general physical qubit errors.
%The pathway for a physical implementation using Rydberg atoms is also novel, as well as the presented quantum Monte-Carlo simulations of the qubot dynamics and the inclusion of effects due to contact of the device with a thermal bath}. 

From where we stand, several directions for future exploration can be sighted. For instance, a more thorough investigation of the capabilities of the proposed \textit{entanglement qubot} remains to be done: by tuning the relevant parameters such as the Rydberg level detunings $ \Delta_{\pm} $ and trap potential $ V_{t}(R) $ can we engineer a qubot capable of protecting entangled states other than the $ \vert \phi^{+} \rangle $ state? What about implementing a system analogous to the conceptual model, capable of protecting an arbitrary logical qubit? Could we extend the device to handle multiple qubits? Would the protection work against general physical errors?
We have focused on the implementation using Rydberd-dressed atoms, but that is certainly not the only possibility. What other opportunities are offered by considering different physical setups for qubots? Polar molecules provide a promising platform \cite{Wei2011, Micheli2006, Brennen2007, Carr2009} with the possibility of coupling to superconducting quantum electronics \cite{Andre2006}. 

% and quantum mechanical motion coupled to internal spin degrees of freedom. 
Synthetic molecular machines are one of the frontiers of nanotechnology \cite{Zhang2018, Stoddart2015, Kassem2017, Lau2017}. Enabled by the idea of a quantum robot we can envision extensions of the molecular machinery toolbox where the quantum states of the nanomachines play a fundamental role in their dynamics.
These devices would combine resources from the environment, stochasticity and non-equilibrium to execute coupled quantum motion and processing of quantum information entering the realm of quantum nanomechanics. For example, in the entanglement qubot one could set the correction sites to perform the operation $ L1 = L2 = X_{b} $, and initiate the spins in the state $ \vert \psi^{+} \rangle $. This would cause a periodic spin-driven motion of the atom. It would be interesting to investigate the possibility of building quantum time crystals \cite{Choi2017, Wilczek2012, Zhang2017} using this scheme.

Quantum robots with no moving parts are also a hitherto unexplored direction. In such devices an error in one degree of freedom would unleash a chain of reactions in other internal non-mechanical parts of the system, which would act back on the affected degree of freedom and steer it to a desired state. This touches upon the theoretical issue of quantum feedback \cite{Milburn, Ahn2002}, in a situation where the feedback itself is carried by quantum mechanical information, rather than the usual classical information scheme in which a measurement result is used to counter-act on the system.

Finally, a very intriguing thought is the combination of a large number of quantum robots interacting with each other. Large numbers of interacting classical \textit{active} agents display fascinating emergent behavior \cite{Vicsek1995, Geyer2018}. Ensembles of active quantum agents on the other hand remain unexplored. Qubots offer a concrete path towards experimentally uncovering the physics of \textit{quantum active} matter.

\section*{Acknowledgements}
I thank Bruno Melo, Lucianno Defaveri, Bruno Suassuna, Igor Brand\~ao and George Svetlichny for useful discussions and feedback on the manuscript. This work was financed in part by the Coordenac\~ao de Aperfei\c{c}oamento de Pessoal de N\'ivel Superior - Brasil (CAPES) - Finance Code 001, Conselho Nacional de Desenvolvimento Cient\'ifico e Tecnol\'ogico (CNPq) and the FAPERJ Scholarship No. E-26/202.830/2019.
\\

%To be discussed here:
%
%\begin{itemize}
%
%\item Capabilities of this qubot; different potentials, stabilization of different states.
%
%\item Full qubot.
%
%\item Alternative systems: molecules, optomechanics in motion.
%
%\item Possibility of a qubot with no moving parts.
%
%\item Ensembles of qubots: active quantum matter?
%
%\end{itemize}

\pagebreak
\onecolumngrid
\section{Appendix A: effective potentials}

As described in the main text, admixing strongly interacting Rydberg states from the $ n^{2}P_{1/2} $ manifold to the low-lying $ 5^{2}S_{1/2} $ Zeeman sublevels induces spatial dependent spin-spin interactions of the form \eqref{real_int}. 
For completeness we reproduce the main results of \cite{Glaetzle2015} outlining the toolbox for engineering a wide range of effective spin interactions.

The interaction coefficients $ J_{\alpha} $ are calculated by adiabatic elimination of the Rydberg levels $ \vert r_{\pm} \rangle $ up to fourth order in $ \Delta / \Omega $, and are given by
\begin{align}
J_{z}(R) = \dfrac{1}{4} \left(  \tilde{V}_{--}(R) - 2 \tilde{V}_{+-}(R)  + \tilde{V}_{++}(R) \right) \ , \\
J_{x}(R) = 2 \left(  \tilde{W}_{+-}(R) +  \tilde{W}_{++}(R) \right)  \ , \\
J_{y}(R) = 2 \left(  \tilde{W}_{+-}(R) -  \tilde{W}_{++}(R) \right) \ , \\
J_{\parallel}(R) = \dfrac{1}{4} \left(  \tilde{V}_{--}(R) - \tilde{V}_{++}(R) \right) \ ,
\end{align}
where the functions $ \tilde{W}_{\alpha \beta}, \tilde{V}_{\alpha \beta} $ are effective radial dependent steplike potentials,
%\begin{widetext}
\begin{eqnarray}
\tilde{V}_{\alpha \alpha}(R) = \dfrac{\Omega^{2}_{\bar{\sigma}}}{2 \Delta_{\bar{\sigma}}} - \dfrac{\Omega^{4}_{\bar{\sigma}}}{4 \Delta_{\bar{\sigma}}^{3}} + \dfrac{\Omega_{\bar{\alpha}}^{4}}{4 \Delta_{\bar{\alpha}}^{2}} \dfrac{  V_{++} - 2 \Delta_{\alpha} }{W_{++}^{2} - ( V_{++} - 2 \Delta_{+} )( V_{++} - 2 \Delta_{-} ) } 
\end{eqnarray}
\begin{eqnarray}
\tilde{V}_{+-}(R) = \dfrac{\Omega^{2}_{-}}{4 \Delta_{-}} + \dfrac{\Omega^{2}_{+}}{4 \Delta_{+}} - \dfrac{\Omega_{+}^{2} \Omega_{-}^{2}}{16 \Delta_{+}^{2} \Delta_{-}} -   \dfrac{\Omega_{+}^{2} \Omega_{-}^{2}}{16 \Delta_{-}^{2} \Delta_{+}}  - \dfrac{\Omega^{4}_{-}}{16 \Delta_{-}^{3}} - \dfrac{\Omega^{4}_{+}}{16 \Delta_{+}^{3}}  + \dfrac{\Delta_{\pm}^{2}  \Omega_{+}^{2} \Omega_{-}^{2}}{16 \Delta_{+}^{2} \Delta_{-}^{2}} \dfrac{ ( \Delta_\pm - V_{+-} ) }{(\Delta_\pm - V_{+-})^{2} - W_{+-}^{2}}  
\label{steplike1}
\end{eqnarray}
\ 
\begin{eqnarray}
\tilde{W}_{+-}(R) &=& \dfrac{\Omega_{+}^{2} \Omega_{-}^{2}}{16 \Delta_{+}^{2} \Delta_{-}^{2}}   \dfrac{ \Delta_{\pm}^{2} W_{+-}  }{ (\Delta_\pm - V_{+-})^{2} - W_{+-}^{2}  } 
\end{eqnarray}
\begin{eqnarray}
\tilde{W}_{++}(R) &=& \dfrac{\Omega_{+}^{2} \Omega_{-}^{2}}{4 \Delta_{+} \Delta_{-}} \dfrac{W_{++}}{W_{++}^{2} - ( V_{++} - 2 \Delta_{+} )( V_{++} - 2 \Delta_{-} ) } 
 \label{steplike2}
\end{eqnarray}
%\end{widetext}
written in terms of the $ n^{2}P_{1/2} $ van der Waals potentials $ V_{\alpha \beta}, W_{\alpha \beta} $. Note the single particle light-shifts have been included in the above expressions.
Moreoever, $ \tilde{V}_{+-} = \tilde{V}_{-+} $,  and we have defined $ \Delta_{\pm} = \Delta_{+} + \Delta_{-} $ and $ \bar{\alpha} = - \alpha $. In the parameter region $ \Delta_{+-} < 0 $, $ \Delta_{+} / \Delta_{-} < 0 $ resonant Rydberg excitations are avoided for all values of $ R $.  
For atomic orientation $ \theta = \pi / 2 $ (polar), $ \phi = 0 $ (azimuthal) the  van der Waals potentials are 
\begin{eqnarray}
V_{\alpha \beta} = \dfrac{c_{\alpha \beta}}{R^{6}} \ , \ W_{+-}  = \dfrac{w}{R^{6}} = -\dfrac{1}{3} W_{++} \ .
\end{eqnarray}
where the so-called $ C_{6} $ coefficients $ c_{\alpha \beta} $ and $ w $ are obtained from second order perturbation theory, and are given by
\begin{eqnarray}
c_{++} &=&\dfrac{2}{81}  \left( 5  C_{6}^{(a)} + 14  C_{6}^{(b)}  + 8  C_{6}^{(c)} \dfrac{}{}  \right) \\
c_{+-} &=&\dfrac{2}{81}  \left(   C_{6}^{(a)} + 10  C_{6}^{(b)}  + 16  C_{6}^{(c)} \dfrac{}{}  \right) \\
w &=&\dfrac{2}{81}  \left(  C_{6}^{(a)} +   C_{6}^{(b)}  - 2  C_{6}^{(c)} \dfrac{}{}  \right)
\end{eqnarray}
The \textit{indivitual channel} coefficients $ C_{6}^{(\nu)} $, $ \nu = a,b,c $ are not dependent of magnetic quantum numbers and characterize the interaction strengh. 
There is one channel for each non-vanishing matrix element of the dipole-dipole interaction potential \cite{Glaetzle2015}, 
\begin{eqnarray}
a &:& \ P_{1/2} + P_{1/2} \rightarrow S_{1/2} + S_{1/2} \\
b &:& \ P_{1/2} + P_{1/2} \rightarrow D_{3/2} + D_{3/2} \\
c &:& \ P_{1/2} + P_{1/2} \rightarrow D_{3/2} + S_{1/2}
\end{eqnarray}
and each $ C_{6}^{(\nu)} $ is calculated from the radial part of the dipole-dipole matrix element \cite{Walker2007},
\begin{eqnarray}
C^{(\nu)}_{6} = \sum_{n_{\alpha} n_{\beta}} \dfrac{e^{4}}{\delta_{\alpha \beta}} \left(  R_{nl}^{n_{\alpha} l_{\alpha}}  R_{nl}^{n_{\beta} l_{\beta}}  \right)^{2}
\label{C6_coeff}
\end{eqnarray}
where
\begin{eqnarray}
R_{nl}^{n_{i} l_{i}}  = \int dr r^{2} \psi_{n,l,j}(r)^{*} r \psi_{n_{i},l_{i},j_{i}}(r) \ ,
\end{eqnarray}
and $ \delta_{\alpha \beta} $is the energy defect between levels $  n_{\alpha}$ and $ n_{\beta} $. 

\begin{figure}[ht!] % [width=8.6 cm]
%    \centering
    \includegraphics[width =0.5\textwidth]{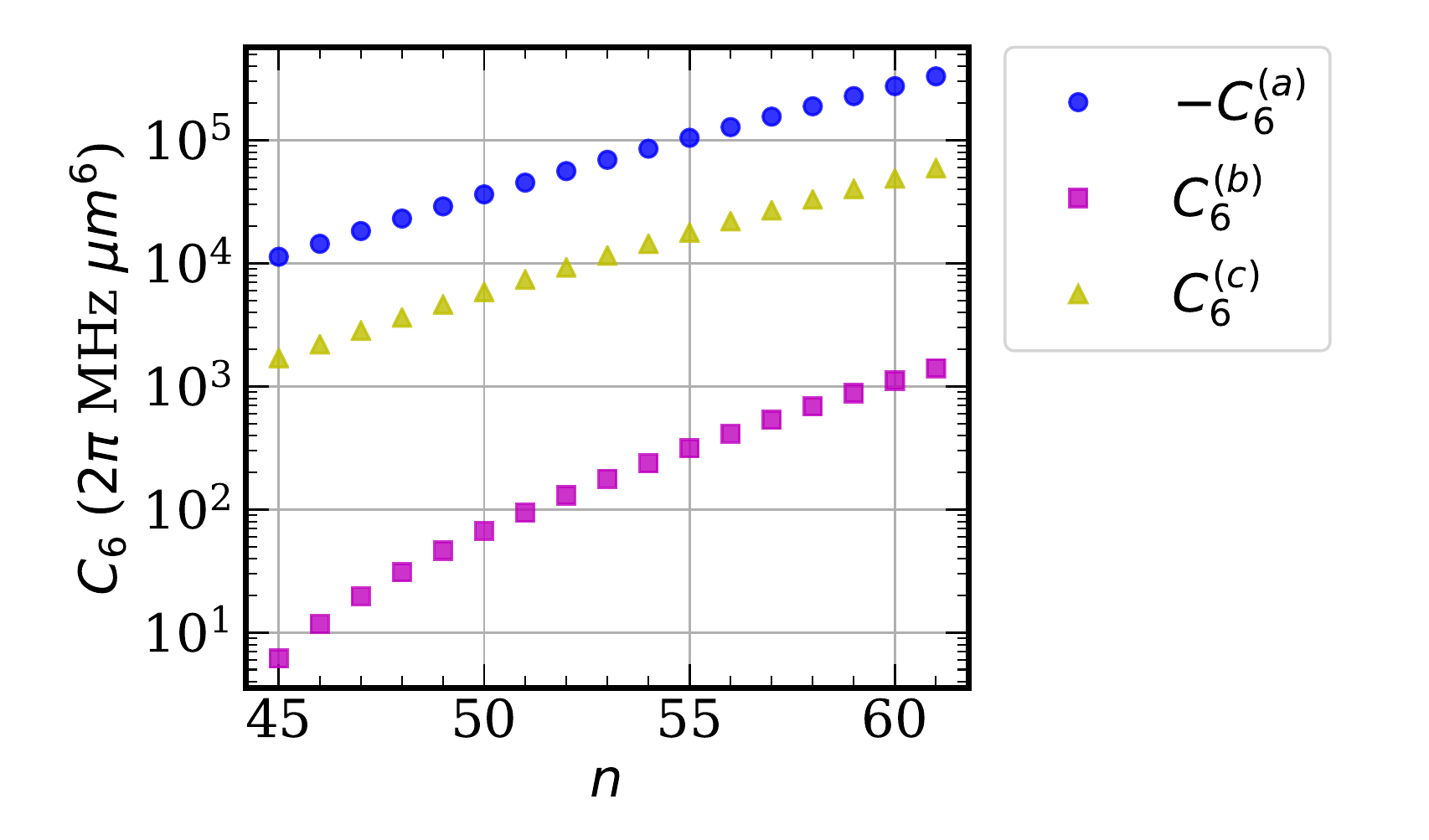}
    \caption[]{$ C_{6}^{(\nu)} $ coefficients as a function of principal quantum number for the $ n^{2}P_{1/2} $ manifold. }\label{C6}
    \label{color}
\end{figure}

To numerically obtain the coefficients \eqref{C6_coeff}, and consequently the step-like potentials \eqref{steplike1} and \eqref{steplike2}, we use the ARC python library for alkali Rydberg atoms \cite{ARC}. Numerical calculation results are shown in Figure \ref{C6} as a function of the principal quantum number for the $ n^{2}P_{1/2} $ manifold. For $ n = 60 $, as used in the main text, we find
\begin{eqnarray}
- C_{6}^{(a)} \approx 2 \pi \times \SI{2.7E5}{MHz \cdot \mu m^{6}} \\
C_{6}^{(b)} \approx 2 \pi \times \SI{1.1E3}{MHz \cdot \mu m^{6}} \\
C_{6}^{(c)} \approx  2 \pi \times \SI{4.9E4}{MHz \cdot \mu m^{6}}
\end{eqnarray}

\section{Appendix B: Magnetic field $ J_{\parallel} $}

\begin{figure}[ht!] % [width=8.6 cm]
%    \centering
    \includegraphics[width =0.45\textwidth]{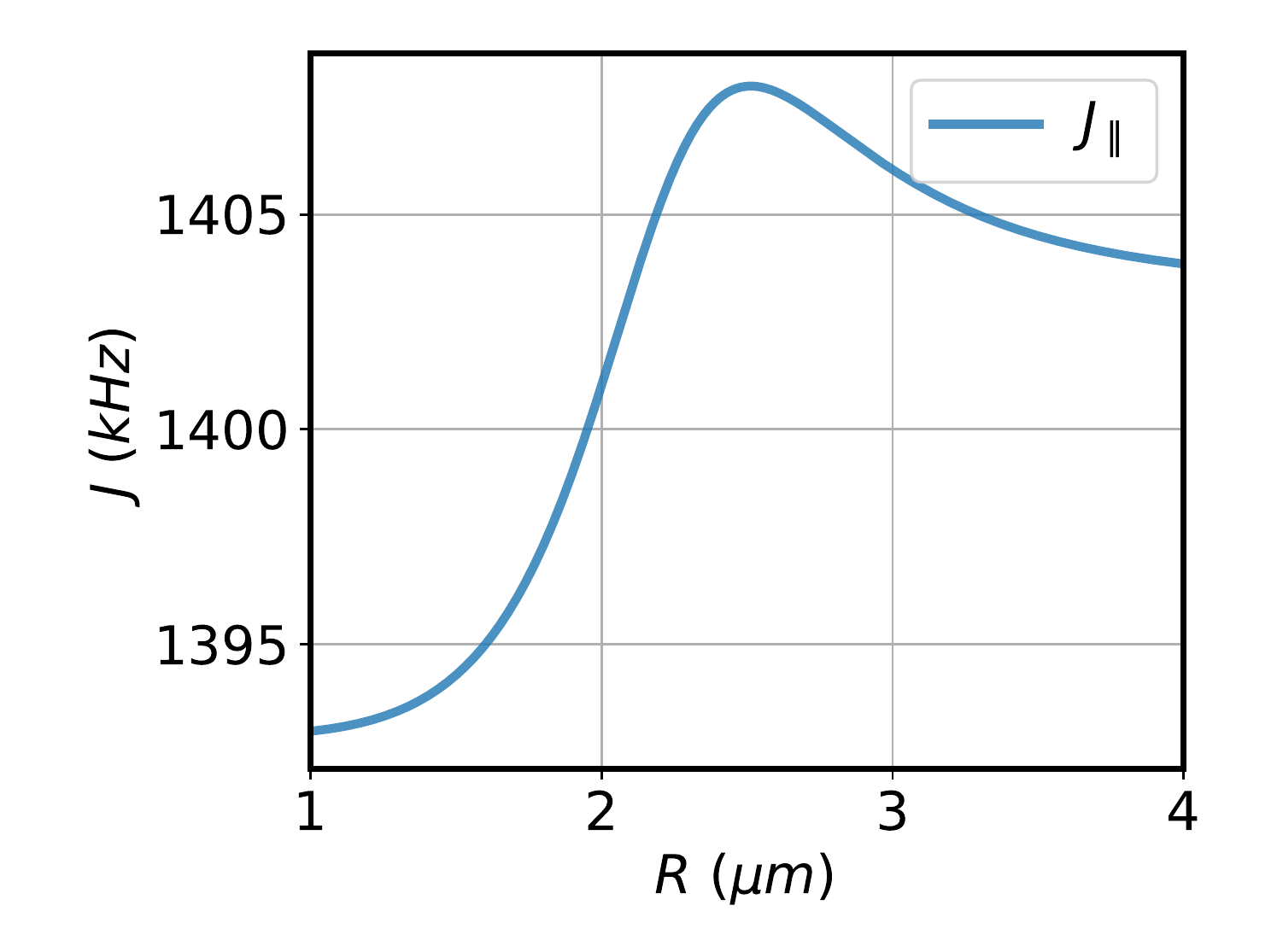}
    \caption[]{$ J_{\parallel} $ profile. }\label{J_par}
    \label{color}
\end{figure}

Besides the $ J_{\alpha}(R) $ coefficients, the Rydberg dressing generates an effective magnetic field term $ J_{\parallel} (Z_{a} + Z_{b}) $ in the interaction energy. Under the influence of this term, Bell states of the $ ab $ pair are no longer eigenstates of the interaction. To obtain the spin dependent potential landscapes given by the eigenvalues in Eqs.\eqref{V_1}-\eqref{V_4}, we need to cancel $ J_{\parallel} $ by applying an external spatial dependent static field.
How large such a field needs to be?
A plot of $ J_{\parallel} $ can be seen in Figure \ref{J_par}. 

Note that $ \langle J_{\parallel} \rangle \approx \SI{1401}{kHz} $. Considering the Land\'e factor $ \vert g_{F} \vert \approx \SI{0.70}{MHz / G} $ for the $ 5^{2} S_{1/2} $ states \cite{Bize1999} this effective magnetic field can be cancelled by an additional weak non-homogeneous field of order of magnitude $ \vert B_{c} \vert \approx \SI{2}{G} $.

\section{Appendix C: Alternative spin pattern}

\begin{figure}[ht!] % [width=8.6 cm]
    \centering
    \includegraphics[width =\textwidth]{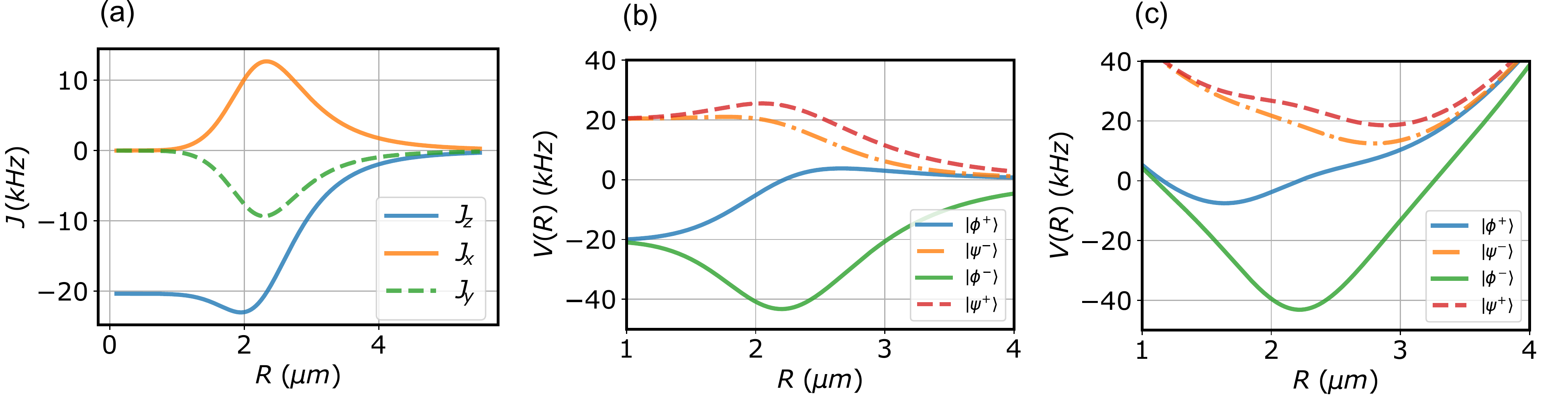}
    \caption[]{(a) Alternative spin pattern profile. (b) Resulting spin dependent potentials. (c) Resulting spin dependent potentials after adding the spin-independent harmonic potential. }\label{alternative_potentials}
    \label{color}
\end{figure}

\begin{figure}[ht!] % [width=8.6 cm]
%    \centering
    \includegraphics[width =0.45\textwidth]{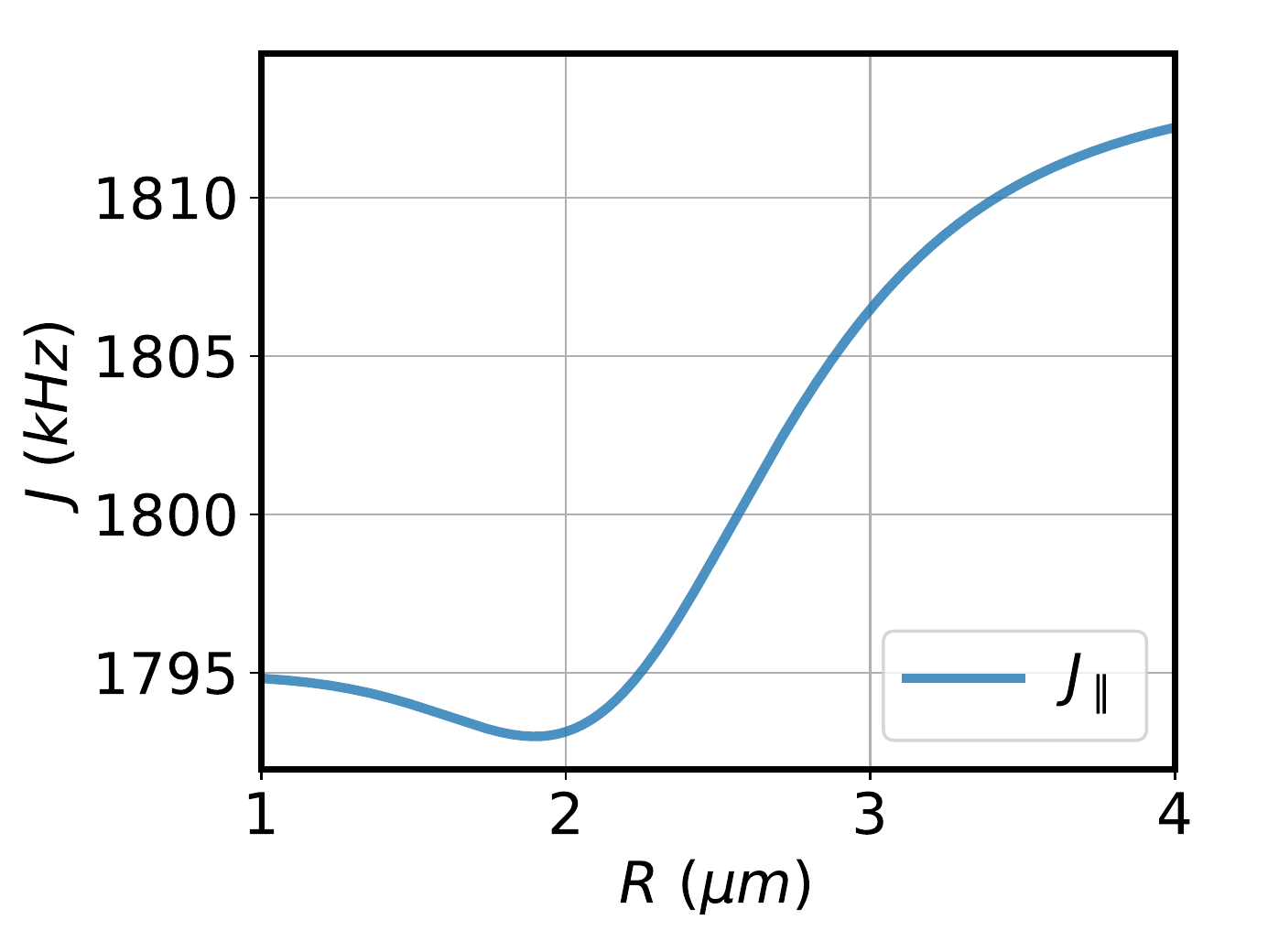}
    \caption[]{Alternative effective magnetic field. }\label{alternative_J_par}
    \label{color}
\end{figure}

Alternative spin dependent potentials, defined by parameters different from the ones employed in the main text are shown in Figure \ref{alternative_potentials}. Here, we consider detunings $ \Delta_{+} = -2 \pi \times \SI{70}{MHz} $, $ \Delta_{-} = 2 \pi \times \SI{30}{MHz}  $, Rabi frequencies $ \Omega_{+} = \Omega_{-} = -2 \pi \times \SI{7}{MHz} $ and the trap potential
\begin{eqnarray}
V_{t}(R) = V_{0} \left( R - \delta   \right)^{2}
\end{eqnarray}
where $ V_{0} = \SI{15}{kHz / \mu m^{2}} $ and $ \delta = \SI{2.30}{\mu m} $.
Note the resulting landscapes in Figure \ref{alternative_potentials}(c) suggest $ \vert \phi^{-} \rangle $ as protected state, while corrective loops $ L1$ and $ L2 $ should be reversed with respect to the choice discussed in the main text.
The effective magnetic field has a mean value $ \langle J_{\parallel} \rangle \approx \SI{1803}{kHz} $, which requires a slightly higher compensating magnetic field, but still on the order of a few Gauss. The spatial profile $ J_{\parallel}(R) $ is shown in Figure \ref{alternative_J_par}. 

\section{Appendix D: Optimal operation}

To evaluate the effect of positioning of the correctors $ L1 $ and $ L2 $, we ran \textbf{SMMC}, as described in the main text, for different values of the positions $R_{L1} = - R_{L2} $.

\begin{figure}[ht!] % [width=8.6 cm]
%    \centering
    \includegraphics[width =0.5\textwidth]{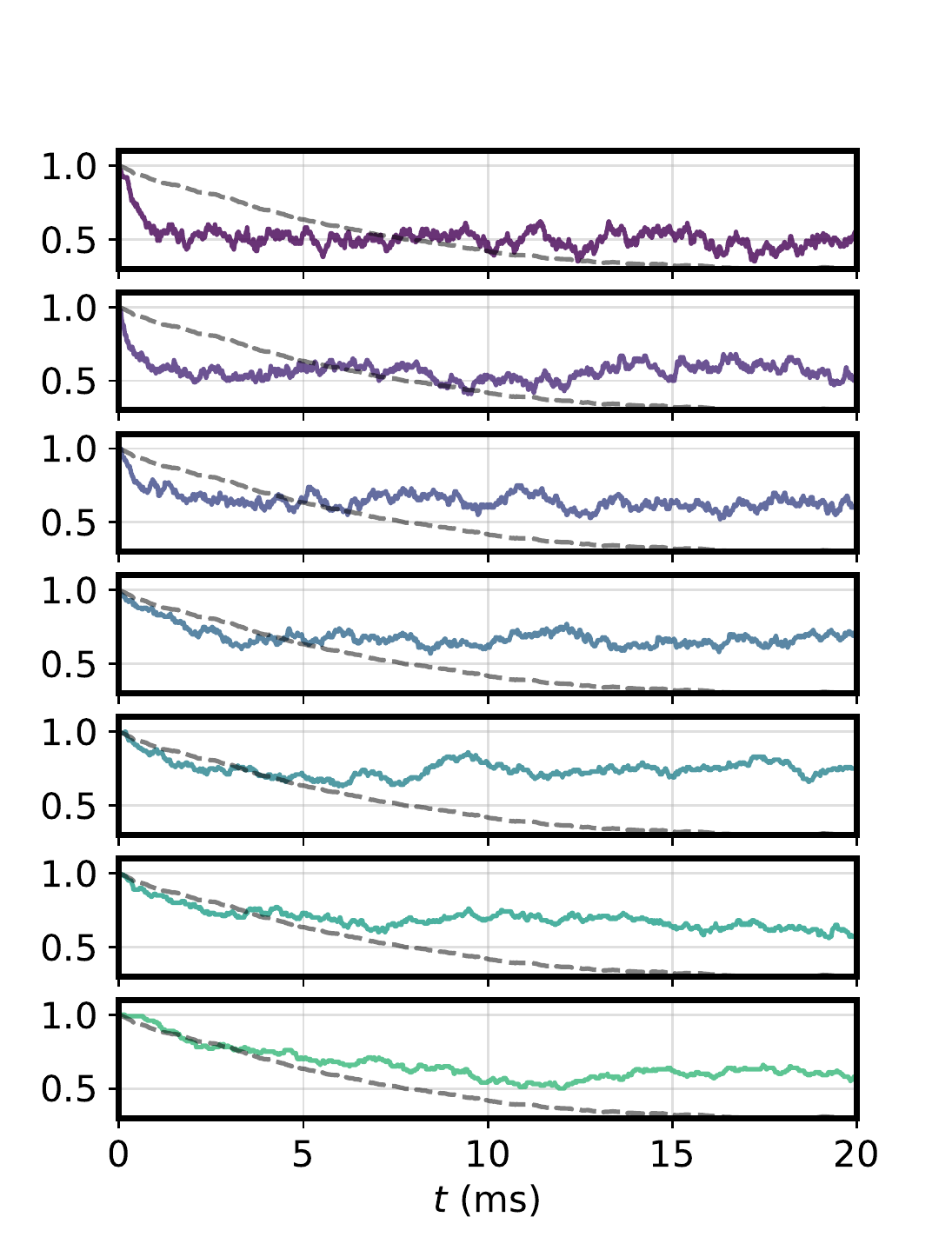}
    \caption[]{ Overlap $ F $ for different values of the corrector position $ \vert R_{L1} \vert = \vert R_{L2} \vert $ obtained from simulating $ 10^{2} $ quantum trajectories. For each trace position values are $ (0.47, 0.52, 0.58, 0.63, 0.69, 0.74, 0.80) \ \SI{}{\mu m}$ from top to bottom, respectively. Grey dashed lines represent the overlap of free decohering spins, for comparison. All remaining parameters are the same as in Figure \ref{simulation1} in the main text. }\label{fidelity_optimal_comparison}
    \label{color}
\end{figure}

Figure \ref{fidelity_optimal_comparison} shows traces of the overlap $ F $ as a function of time. Each trace corresponds to a different corrector position (see caption), and the overlap of free spins under the action of the depolarizing channel is shown as the grey dashed line for comparison.
The points in Figure \ref{optimal} (see main text) are obtained by time-averaging the overlap above $ \SI{10}{ms} $ for each of the traces in Figure \ref{fidelity_optimal_comparison}. 

We can see that if the corrector's positions are too close to the atom's equilibrium position, the overlap quickly decays due to \textit{internal} errors, occurring when a quantum fluctuation in the atomic position places it near the corrective site. This fast drop in overlap can be mitigated by positioning the correctors further apart from the $ \vert \phi^{+} \rangle $ equilibrium point. There is, however, a trade-off: the maximum steady state overlap $ \approx 70 \% $ is reached for a position $ \vert R_{L1} \vert \approx \SI{0.63}{\mu m} $, while placing the correctors further than that reduce the correction rates below the decoherence rate and consequently the steady state overlap. 

Decoherence causes the overlap to decrease exponentially according to $ e^{-\Gamma t} = e^{-t / \tau_D} $, where $ \tau_{D} = \Gamma^{-1} = (\SI{100}{Hz})^{-1} =  \SI{10}{ms} $ is the characteristic decay time of the system. Decoherence effectivelly freezes when the system reaches its steady state, which happens after a stabilization time $ t_s $ elapses. From Figure \ref{simulation1} in the main text, we see that $ t_{s} \approx \SI{4}{ms} $, yielding an expected overlap of $ F \approx e^{-t_{s}/\tau_{D}} \approx 0.67 $, in accordance to the simulation results.
\\

\twocolumngrid

%2ca02ccc
%1f77b4cc

\end{document}